\documentclass[%
 reprint,
 amsmath,amssymb,
 aps, prx
]{revtex4-2}

\usepackage{graphicx}
\usepackage{dcolumn}
\usepackage{bm}
\usepackage{booktabs} 

\usepackage{braket}
\usepackage{multirow}
\usepackage{upgreek}
\usepackage{makecell}
\usepackage{hyperref}
\usepackage{orcidlink}
\hypersetup{colorlinks,linkcolor=blue,anchorcolor=blue,citecolor=blue,urlcolor=blue}

\usepackage{amsmath}
\allowdisplaybreaks

\usepackage{xcolor}

\begin{document}



\title{\textbf{A Concatenated Dual Displacement Code for Continuous-Variable Quantum Error Correction}}%

\author{Fucheng Guo\,\orcidlink{0009-0003-0631-7404}}
\email{fguo22@ncsu.edu}
\affiliation{Department of Computer Science, North Carolina State University, Raleigh, North Carolina 27695-8206, USA
}%

\author{Frank Mueller\,\orcidlink{0000-0002-0258-0294}}
\email{fmuelle@ncsu.edu}
\affiliation{Department of Computer Science, North Carolina State University, Raleigh, North Carolina 27695-8206, USA
}%

\author{Yuan Liu\,\orcidlink{0000-0003-1468-942X}}
\email{q\_yuanliu@ncsu.edu}
\affiliation{Department of Electrical and Computer Engineering, North Carolina State University, Raleigh, North Carolina 27695, USA}%
\affiliation{Department of Computer Science, North Carolina State University, Raleigh, North Carolina 27695, USA}
\affiliation{Department of Physics, North Carolina State University, Raleigh, North Carolina 27695, USA
}%



\date{\today}

\begin{abstract}
The continuous-variable (CV) Gaussian no-go theorem fundamentally
limits the suppression of Gaussian displacement errors using only
Gaussian gates and states. Prior studies have employed
Gottesman-Kitaev-Preskill (GKP) states as ancillary qumodes to
suppress small Gaussian displacement errors. However, when the
displacement magnitude becomes large, inevitable lattice-crossing
errors arise beyond the correctable range of the GKP state. To
address this issue, we concatenate the Gaussian-noise-suppression
circuit with an outer analog Steane code that corrects such occasional
lattice-crossing events as well as other abrupt displacement errors.
Contrary to conventional concatenation, which primarily aims to
reduce logical error rates, the Steane--GKP duality in encoding
provides complementary protection against displacement errors at
different scales: The inner GKP layer employs non-Gaussian resources
to suppress continuous Gaussian noise and reduce residual variance,
while the outer analog Steane code corrects discrete lattice-crossing
events that exceed the GKP correctable range. It is precisely this
separation of error-mitigation roles that enables CV error correction.
In contrast to prior work on concatenating GKP and repetition codes
to establish error correction for discrete qubit/qudit encoding, we
provide correction in the continuous encoding space.
Analytical studies show that, under infinite squeezing, the
concatenated code suppresses the variance of Gaussian displacement
errors acting on all qumodes by up to 50\%, while enabling unbiased
correction of lattice-crossing errors with a success probability
determined by the ratio between the residual Gaussian error standard
deviation and the lattice-crossing magnitude. Even with finite
squeezing, the proposed architecture still provides Gaussian-error
suppression and lattice-crossing correction. Moreover, the presence
of the outer analog Steane code relaxes the squeezing requirement of
the inner GKP states, indicating near-term experimental feasibility.
This work establishes a viable route toward fault-tolerant
continuous-variable quantum computation and provides new insight
into the design of concatenated CV error-correcting architectures.
\end{abstract}

\maketitle


\section{\label{sec:introduction}Introduction}

Quantum error correction (QEC) lies at the foundation of
fault-tolerant quantum computation, enabling reliable information
processing in the presence of noise and
decoherence~\cite{Shor1995,Steane1996,Gottesman1997}.  In addition to
qubit-based schemes, bosonic quantum error correction encodes logical
information into CV modes of harmonic oscillators, providing a
hardware-efficient means to protect against common bosonic errors such
as photon loss and displacement errors~\cite{Cai2021}.  While
discrete-variable (DV) qubit systems have achieved remarkable progress
with stabilizer-based
codes~\cite{Fowler2012,Google2023,Xu2023QubitOscillator}, including
advanced concatenation schemes that combine GKP states with DV
repetition or outer stabilizer codes to enhance error correction and
approach fault-tolerant
operation~\cite{Stafford2023Biased,Li2024NoisyAncilla}, such
architectures fundamentally encode logical information in discrete
degrees of freedom and rely on projective syndrome extraction and
digital correction mechanisms. In contrast, CV quantum architectures,
such as optical modes and superconducting resonators, intrinsically
encode information in continuous quadrature variables, leading to
distinct error models dominated by Gaussian displacement noise and
requiring analog error correction strategies beyond conventional
stabilizer formalisms. These CV systems offer an attractive
alternative owing to their larger Hilbert space, natural compatibility
with bosonic hardware, and the possibility of leveraging phase-space
structure for encoding and noise
suppression~\cite{Braunstein2005,Weedbrook2012,Blais2021}.  However,
the dominant errors in CV platforms take the form of Gaussian
displacement noise, which cannot be fully suppressed using Gaussian
operations and states alone due to the fundamental constraint of the
Gaussian no-go theorem~\cite{Niset2009,Terhal2015}. Overcoming this
limitation is essential for realizing scalable, fault-tolerant CV
quantum computation.

A variety of theoretical frameworks have been proposed to suppress
Gaussian displacement noise in continuous-variable systems that
directly encode continuous logical information rather than discrete
qubits~\cite{Xu2023QubitOscillator}. Among these, GKP-type encodings occupy a central
position, embedding quantum information into a lattice structure in
phase space such that small displacements can be detected and
corrected through modular
measurements~\cite{Gottesman2001,Glancy2006}. Their correction
capability, however, is intrinsically bounded: When the displacement
magnitude exceeds half of the lattice spacing, lattice-crossing errors
arise and cause logical misidentification~\cite{Fukui2018,Noh2020}. In
addition, repetition-type encodings have been explored to redundantly
distribute quantum information across multiple bosonic
modes. Published results include several canonical examples, such as
five-~\cite{Braunstein1998} and nine-wave-packet~\cite{Aoki2009}
codes, which illustrate how continuous logical variables can, in
principle, be encoded with redundancy to protect against local
displacement errors. Nevertheless, these repetition-based
constructions remain ineffective against correlated Gaussian noise
acting collectively on all qumodes, as indicated by the Gaussian no-go
theorem~\cite{Niset2009}. Furthermore, they largely remain theoretical
constructs, as explicit realizations of logical operations and
syndrome-extraction mechanisms~\cite{Schuckert2024} within the encoded
Hilbert space have yet to be developed.

In this work, we propose a concatenated CV error-correction
architecture that integrates an outer analog Steane code (in
Sec.~\ref{sec:cv_steane}) with an inner GKP-assisted noise-suppression
circuit (in Sec.~\ref{sec:gaussian_concatenation}). Our dual approach
combines advantages of the two error correction methods. This
``duality'' refers to a separation of error-mitigation roles in
displacement space, where the inner GKP layer employs non-Gaussian
resources to suppress continuous Gaussian displacement noise and
reduce residual variance, while the outer analog Steane code corrects
discrete lattice-crossing events and other abrupt displacement errors
that exceed the GKP correctable range. At the inner layer Gaussian
displacement noise across all qumodes is continuously suppressed, and
at the outer layer lattice-crossing and abrupt displacement errors are
corrected as they occasionally exceed the GKP correction range. Beyond
establishing the concatenated structure, a complete operational
framework is contributed for the analog Steane code by explicitly
formulating its logical operations and syndrome-extraction mechanisms
within the encoded Hilbert space in Sec.~\ref{sec:cv_steane}. Through
theoretical analysis in Sec.~\ref{sec:performance}, the
error-suppression capability of the concatenated code~\cite{Fukui2023}
is evaluated, the lower bound of achievable noise reduction is
derived, and the experimental feasibility of the proposed architecture
is discussed. Meanwhile, Sec.~\ref{sec:performance} presents Monte
Carlo simulation results, which provide a direct illustration of the
performance of the concatenated code.  These results provide important
theoretical implications for developing scalable and fault-tolerant CV
quantum information
processing~\cite{Chamberland2022}. Sec.~\ref{sec:conclusion} concludes
the paper.

\section{Analog Steane Code}
\label{sec:cv_steane}

In this section, an analog version of the Steane code is constructed based
on its DV counterpart~\cite{Steane1996a}. The corresponding
syndrome-extraction circuits are designed, and the implementation of
fundamental logical operations is discussed. It is rigorously shown
that the proposed analog Steane code can correct single-qumode
displacement errors.

\subsection{Encoding circuit}

\begin{figure}[t]
    \centering
    \includegraphics[width=\linewidth]{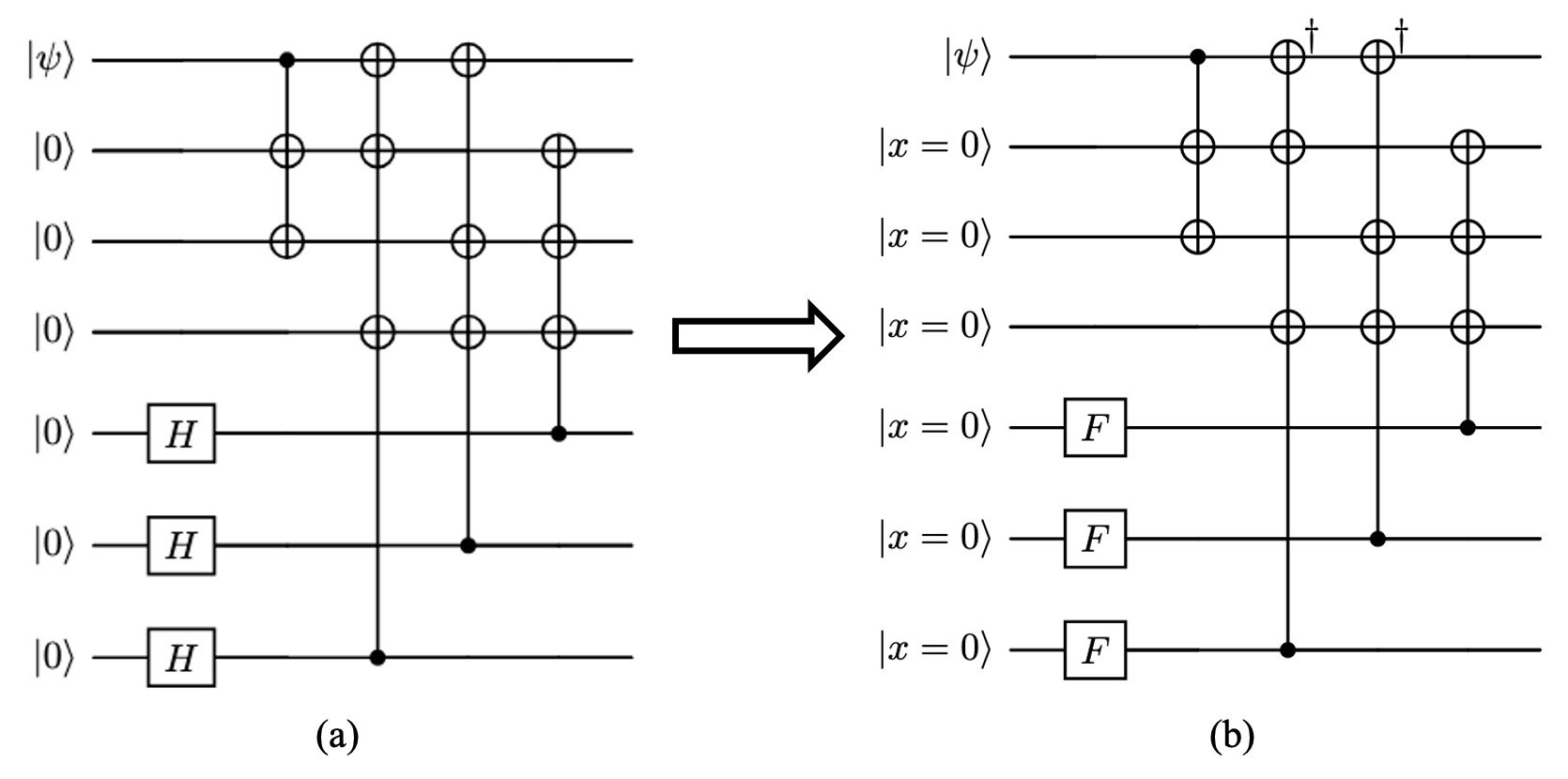}
    \caption{(a) DV Steane code. (b) Analog Steane code obtained by
      replacing the $H$ and CNOT gates with $F$ and SUM (or
      $\mathrm{SUM}^{\dagger}$) gates.}
    \label{fig:Steane}
\end{figure}

First, starting from the DV Steane code shown in
Fig.~\ref{fig:Steane}(a), the Hadamard ($H$) and controlled-NOT (CNOT)
gates are respectively replaced by the Fourier ($F$) gates and the SUM
(or $\mathrm{SUM}^{\dagger}$) gates~\cite{Kalajdzievski2019}. This
substitution yields the CV counterpart of the Steane code, as
illustrated in Fig.~\ref{fig:Steane}(b). Here, The first qumode serves
as the logical qumode, and $\ket{x=0}$ denotes that, in the ideal
case, each qumode is initialized in a position eigenstate, whereas in
practice it corresponds to a finitely squeezed vacuum
state~\cite{Su2018}.

To analyze the structure of the encoded state, the Fourier gate in the
position-quadrature representation is expressed as~\cite{Liu2025}

\begin{eqnarray}
\hat{\mathcal{F}}\ket{x}
  &=& \frac{1}{\sqrt{\pi}} 
      \int dy\, e^{2ixy}\ket{y},
\label{eq:Fourier_x}
\end{eqnarray}
where both $x$ and $y$ are variables in the position basis.  Only the
encoding structure in the position quadrature is analyzed, since the
momentum quadrature is conjugate to the position one.  Therefore, the
corresponding structure in the momentum quadrature can be readily
obtained via a Fourier transformation and is not discussed further in
this paper.

Similarly, the SUM gate in the position quadrature can be expressed
as~\cite{Yoshikawa2008}
\begin{subequations}
\label{eq:SUM_pair}
\begin{eqnarray}
\mathrm{SUM}\ket{x_1, x_2}
  &=& \ket{x_1,\, x_1 + x_2},
\label{eq:SUM_a}
\end{eqnarray}
\begin{eqnarray}
\mathrm{SUM}^{\dagger}\ket{x_1, x_2}
  &=& \ket{x_1,\, x_2 - x_1},
\label{eq:SUM_b}
\end{eqnarray}
\end{subequations}
where both $x_1$ and $x_2$ are variables in the position basis. 
The variable $x_1$ corresponds to the control qumode, while $x_2$
represents the target qumode. Based on the action of the Fourier and
SUM gates, the structure of the encoded state in the position
quadrature can be obtained as
\begin{eqnarray}
\ket{x_{\mathrm{encoding}}}
  &=& \frac{1}{\pi^{1.5}}
      \int dw\, dy\, dz\,
      \ket{x - y - z}\ket{x + w + z} \nonumber\\
  && \quad \times\,
      \ket{x + w + y}\ket{w + y + z}\ket{w}\ket{y}\ket{z},
\label{eq:x_encoding}
\end{eqnarray}
where $x$ represents the encoded logical information, while $w$, $y$,
and $z$ are new variables in the position basis generated through the
action of the Fourier gates.

To demonstrate that the proposed encoding structure possesses
error-correcting capability, it is necessary to show that it satisfies
the Knill-Laflamme condition~\cite{Knill1997,Lloyd1998}, i.e.,
\begin{eqnarray}
\bra{x'_{\mathrm{encoding}}}
\hat{\mathcal{E}}_{\alpha}^{\dagger}\hat{\mathcal{E}}_{\beta}
\ket{x_{\mathrm{encoding}}}
  &=& \delta(x' - x)\,\lambda_{\alpha\beta}, 
  \quad \forall\, \alpha, \beta
\label{eq:KL_condition}
\end{eqnarray}
where $\ket{x_{\mathrm{encoding}}}$ and $\ket{x'_{\mathrm{encoding}}}$
denote two different encoded states,
while $\hat{\mathcal{E}}_{\beta}$ represents a correctable displacement error 
acting on the $\beta$th qumode. 
The coefficient $\lambda_{\alpha\beta}$ is a complex constant
independent of the encoded states.  This condition indicates that
correctable errors do not affect the orthogonality between distinct
encoded states.  For example, when errors occur on qumode~1 and
qumode~2 in the two encoded subspaces, we have
\begin{align}
&\bra{x'_{\mathrm{encoding}}}
  \hat{\mathcal{E}}_{1}^{\dagger}\hat{\mathcal{E}}_{2}
  \ket{x_{\mathrm{encoding}}}
\notag\\
&= \frac{1}{\pi^{3}}
  \int dw'\, dy'\, dz'\, dw\, dy\, dz\,
  \bra{x' - y' - z'} \hat{\mathcal{E}}_{1}^{\dagger} \ket{x - y - z}
\notag\\
&\quad\times
  \bra{x' + w' + z'} \hat{\mathcal{E}}_{2} \ket{x + w + z}
\notag\\
&\quad\times
  \delta(x' + w' + y' - x - w - y)
\notag\\
&\quad\times
  \delta(w' + y' + z' - w - y - z)
\notag\\
&\quad\times
  \delta(w' - w)\,
  \delta(y' - y)\,
  \delta(z' - z)
\notag\\
&= \frac{1}{\pi^{3}}\,
  \delta(x' - x)\,
  \bra{x'} \hat{\mathcal{E}}_{1}^{\dagger} \ket{x}\,
  \bra{x'} \hat{\mathcal{E}}_{2} \ket{x}.
\label{eq:KL_example}
\end{align}

It can be seen that the above expression is nonzero if and only if
$x = x'$.  This indicates that the present encoding scheme can
distinguish errors occurring on qumode~1 and qumode~2.  Similarly, it
can be shown that errors acting on any pair of distinct qumodes within
the encoded space are distinguishable.

By expressing the encoded space in terms of the position and momentum
operators, we can express it as a system of equation per qumode
\begin{align}
\text{qumode1:}\quad 
& \hat{x}_{1}^{(\mathrm{enc})} = -\hat{p}_{6} - \hat{p}_{7} + \hat{x}_{1}, \nonumber\\
& \hat{p}_{1}^{(\mathrm{enc})} = \hat{p}_{1} - \hat{p}_{2} - \hat{p}_{3}, \nonumber\\[2pt]
\text{qumode2:}\quad 
& \hat{x}_{2}^{(\mathrm{enc})} = \hat{p}_{5} + \hat{p}_{7} + \hat{x}_{1} + \hat{x}_{2}, \nonumber\\
& \hat{p}_{2}^{(\mathrm{enc})} = \hat{p}_{2}, \nonumber\\[2pt]
\text{qumode3:}\quad 
& \hat{x}_{3}^{(\mathrm{enc})} = \hat{p}_{5} + \hat{p}_{6} + \hat{x}_{1} + \hat{x}_{3}, \nonumber\\
& \hat{p}_{3}^{(\mathrm{enc})} = \hat{p}_{3}, \nonumber\\[2pt]
\text{qumode4:}\quad 
& \hat{x}_{4}^{(\mathrm{enc})} = \hat{p}_{5} + \hat{p}_{6} + \hat{p}_{7} + \hat{x}_{4}, \nonumber\\
& \hat{p}_{4}^{(\mathrm{enc})} = \hat{p}_{4}, \nonumber\\[2pt]
\text{qumode5:}\quad 
& \hat{x}_{5}^{(\mathrm{enc})} = \hat{p}_{5}, \nonumber\\
& \hat{p}_{5}^{(\mathrm{enc})} = -\hat{p}_{2} - \hat{p}_{3} - \hat{p}_{4} - \hat{x}_{5}, \nonumber\\[2pt]
\text{qumode6:}\quad 
& \hat{x}_{6}^{(\mathrm{enc})} = \hat{p}_{6}, \nonumber\\
& \hat{p}_{6}^{(\mathrm{enc})} = \hat{p}_{1} - \hat{p}_{2} - 2\hat{p}_{3} - \hat{p}_{4} - \hat{x}_{6}, \nonumber\\[2pt]
\text{qumode7:}\quad 
& \hat{x}_{7}^{(\mathrm{enc})} = \hat{p}_{7}, \nonumber\\
& \hat{p}_{7}^{(\mathrm{enc})} = \hat{p}_{1} - 2\hat{p}_{2} - \hat{p}_{3} - \hat{p}_{4} - \hat{x}_{7},
\label{eq:xp_encoding}
\end{align}

where $\hat{x}^{(\mathrm{enc})}$ and $\hat{p}^{(\mathrm{enc})}$ denote
the position and momentum operators after encoding, while all other
operators correspond to those in the initial state. The logical
information is encoded in the first mode, whose quadrature operators
$\hat{x}_1$ and $\hat{p}_1$ represent the logical position and momentum,
respectively.

\subsection{Syndrome Extraction and Error Correction}

As previously demonstrated, the encoding circuit provides sufficient
redundancy to distinguish displacement errors occurring on different
qumodes.  Next, it is necessary to design a circuit that extracts the
syndrome and, based on the measurement outcomes, identifies and
corrects the corresponding errors~\cite{Hockings2025,Welsch1999}.

\begin{figure*}[t]
    \centering
    \includegraphics[width=1\linewidth]{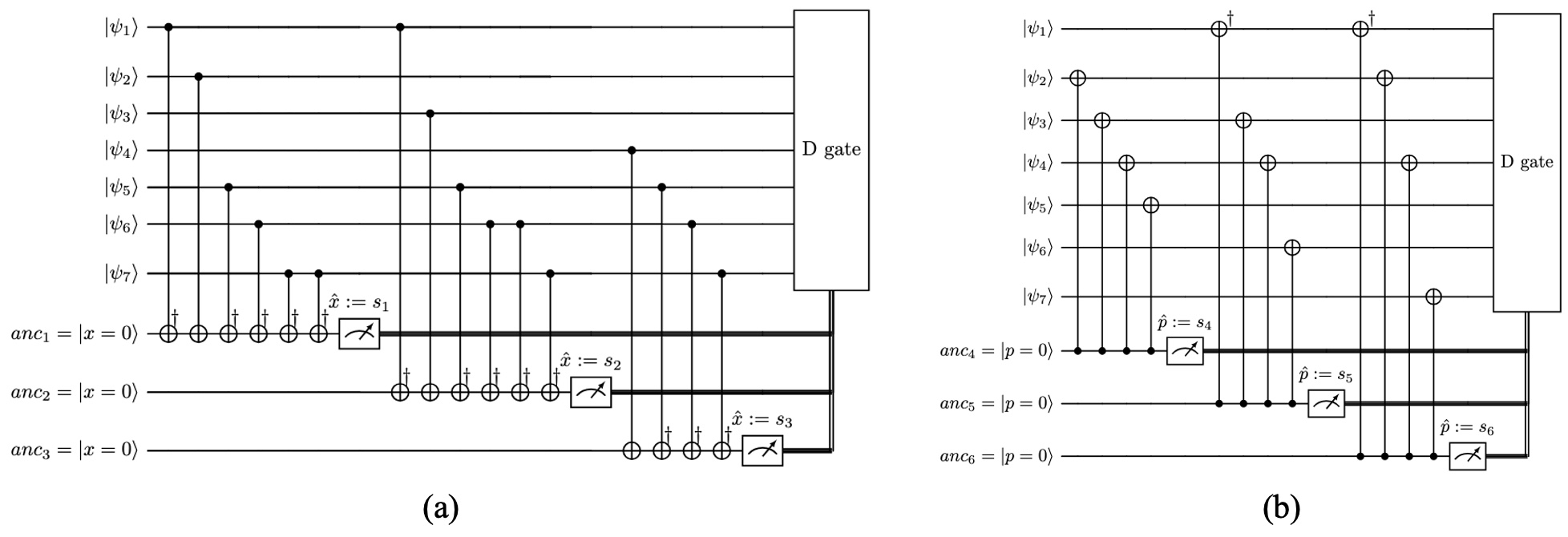}
    \caption{(a) Position-quadrature syndrome extraction circuit. (b)
      Momentum-quadrature syndrome extraction circuit. Both circuits
      employ three ancilla qumodes, each initialized in a position or
      momentum eigenstate, respectively, to facilitate syndrome
      readout. Homodyne measurements are performed on the
      position/momentum quadratures of the ancilla qumodes, yielding
      syndromes $s_1$ $\sim$ $s_6$. By analyzing these syndromes, the
      location and magnitude of the deterministic displacement errors
      can be identified, and the corresponding errors are corrected
      through displacement (D) gates.}
    \label{fig:xsyndrome}
\end{figure*}

The circuit shown in Fig.~\ref{fig:xsyndrome} corresponds to the
position and momentum quadratures syndrome extraction circuits.  For
each circuit, three ancilla qumodes are introduced, each initialized
in a position or momentum eigenstate.  These ancillae are coupled to
the qumodes in the encoded space through a sequence of $\mathrm{SUM}$
or $\mathrm{SUM}^{\dagger}$ gates.  Subsequently, homodyne
measurements are performed on the position/momentum quadratures of the
ancilla qumodes, and the obtained measurement outcomes correspond to
the syndrome values.  Based on the obtained syndrome values, error
localization and magnitude estimation are performed, followed by error
correction implemented through displacement (D) gates.

Let us assume that the errors occurring on each qumode within the encoded
block are given by
\begin{align}
\boldsymbol{\epsilon} =
\begin{bmatrix}
\epsilon_{x1} & \epsilon_{p1} &
\epsilon_{x2} & \epsilon_{p2} &
\cdots &
\epsilon_{x7} & \epsilon_{p7}
\end{bmatrix}^{\top},
\label{eq:error_vector}
\end{align}
where $\epsilon_{x}$ and $\epsilon_{p}$ denote displacement errors
occur in the position and momentum quadratures of a single qumode,
respectively.

According to the syndrome extraction circuit shown in
Fig.~\ref{fig:xsyndrome}, the expressions for the syndromes can be
written as
\begin{equation}
\begin{aligned}
s_1 &= -\epsilon_{x1} + \epsilon_{x2} - \epsilon_{x5} - \epsilon_{x6} - 2\epsilon_{x7} + \hat{x}_2 + \hat{x}_{\mathrm{anc1}},\\
s_2 &= -\epsilon_{x1} + \epsilon_{x3} - \epsilon_{x5} - 2\epsilon_{x6} - \epsilon_{x7} + \hat{x}_3 + \hat{x}_{\mathrm{anc2}},\\
s_3 &= \epsilon_{x4} - \epsilon_{x5} - \epsilon_{x6} - \epsilon_{x7} + \hat{x}_4 + \hat{x}_{\mathrm{anc3}},\\
s_4 &= -\epsilon_{p2} - \epsilon_{p3} - \epsilon_{p4} - \epsilon_{p5} + \hat{x}_5 + \hat{p}_{\mathrm{anc4}},\\
s_5 &= \epsilon_{p1} - \epsilon_{p3} - \epsilon_{p4} - \epsilon_{p6} + \hat{x}_6 + \hat{p}_{\mathrm{anc5}},\\
s_6 &= \epsilon_{p1} - \epsilon_{p2} - \epsilon_{p4} - \epsilon_{p7} + \hat{x}_7 + \hat{p}_{\mathrm{anc6}}.
\end{aligned}
\label{eq:syndrome}
\end{equation}

Under ideal conditions, all qumodes except the one carrying the
logical information are initialized in position/momentum eigenstates.
Consequently, the mean and variance of $\hat{x}$ and $\hat{p}$ vanish,
allowing all position/momentum operators in Eq.~(\ref{eq:syndrome}) to
be neglected~\cite{GonzalezArciniegas2021}.  The syndrome expressions
can therefore be simplified as
\begin{equation}
\begin{aligned}
s_1 &= -\epsilon_{x1} + \epsilon_{x2} - \epsilon_{x5} - \epsilon_{x6} - 2\epsilon_{x7},\\
s_2 &= -\epsilon_{x1} + \epsilon_{x3} - \epsilon_{x5} - 2\epsilon_{x6} - \epsilon_{x7},\\
s_3 &= \epsilon_{x4} - \epsilon_{x5} - \epsilon_{x6} - \epsilon_{x7},\\
s_4 &= -\epsilon_{p2} - \epsilon_{p3} - \epsilon_{p4} - \epsilon_{p5},\\
s_5 &= \epsilon_{p1} - \epsilon_{p3} - \epsilon_{p4} - \epsilon_{p6},\\
s_6 &= \epsilon_{p1} - \epsilon_{p2} - \epsilon_{p4} - \epsilon_{p7}.
\end{aligned}
\label{eq:syndrome_all}
\end{equation}

Assuming that the displacement errors of each qumode have a magnitude
of unity, the corresponding syndrome values are summarized in
Table~\ref{tab:syndrome_values}.  Different error patterns yield
distinct syndrome values, enabling unique identification of the error
location and estimation of its magnitude.  However, the lookup-based
approach is inefficient in practice.  To improve efficiency, a more
direct analytical procedure for error localization and magnitude
estimation is introduced.

\begin{table}[t]
\caption{\label{tab:syndrome_values}%
Syndrome values with the deterministic displacement magnitude $\epsilon$ taken as unity.}
\begin{ruledtabular}
\begin{tabular}{c|ccc|ccc}
\multirow{2}{*}{Error pattern}
 & \multicolumn{3}{c|}{Position syndromes} 
 & \multicolumn{3}{c}{Momentum syndromes} \\
 & $s_1$ & $s_2$ & $s_3$ & $s_4$ & $s_5$ & $s_6$ \\ \hline
$\epsilon_{x1}=1$ & $-1$ & $-1$ & $0$ & $0$ & $0$ & $0$ \\
$\epsilon_{x2}=1$ & $+1$ & $0$ & $0$ & $0$ & $0$ & $0$ \\
$\epsilon_{x3}=1$ & $0$ & $+1$ & $0$ & $0$ & $0$ & $0$ \\
$\epsilon_{x4}=1$ & $0$ & $0$ & $+1$ & $0$ & $0$ & $0$ \\
$\epsilon_{x5}=1$ & $-1$ & $-1$ & $-1$ & $0$ & $0$ & $0$ \\
$\epsilon_{x6}=1$ & $-1$ & $-2$ & $-1$ & $0$ & $0$ & $0$ \\
$\epsilon_{x7}=1$ & $-2$ & $-1$ & $-1$ & $0$ & $0$ & $0$ \\
\hline
$\epsilon_{p1}=1$ & $0$ & $0$ & $0$ & $0$ & $+1$ & $+1$ \\
$\epsilon_{p2}=1$ & $0$ & $0$ & $0$ & $-1$ & $0$ & $-1$ \\
$\epsilon_{p3}=1$ & $0$ & $0$ & $0$ & $-1$ & $-1$ & $0$ \\
$\epsilon_{p4}=1$ & $0$ & $0$ & $0$ & $-1$ & $-1$ & $-1$ \\
$\epsilon_{p5}=1$ & $0$ & $0$ & $0$ & $-1$ & $0$ & $0$ \\
$\epsilon_{p6}=1$ & $0$ & $0$ & $0$ & $0$ & $-1$ & $0$ \\
$\epsilon_{p7}=1$ & $0$ & $0$ & $0$ & $0$ & $0$ & $-1$ \\
\end{tabular}
\end{ruledtabular}
\end{table}

Owing to the intrinsic property of Steane codes, the analyses of the
position and momentum quadratures are decoupled.  Hence, the following
discussion focuses on the error localization and magnitude estimation
in the position quadrature, while the momentum quadrature follows an
analogous procedure and yields similar results, which are omitted for
brevity.  For the position quadrature, according to
Eq.~\ref{eq:syndrome_all}, the syndrome vector can be expressed as
$\mathbf{s}_x = \mathbf{M}_x \boldsymbol{\epsilon}$. Here,
$\mathbf{M}_x$ is
\begin{equation}
\mathbf{M}_x =
\begin{bmatrix}
-1 & 1 & 0 & 0 & -1 & -1 & -2 \\
-1 & 0 & 1 & 0 & -1 & -2 & -1 \\
0  & 0 & 0 & 1 & -1 & -1 & -1
\end{bmatrix}.
\label{eq:Mx}
\end{equation}

Defining the column vector $\mathbf{m}_j$ as the $j$th column of
$\mathbf{M}_x$, the following expression can be written as
\begin{equation}
T_j = \frac{\mathbf{m}_j^{\top} \mathbf{s}_x}{\lVert \mathbf{m}_j \rVert}.
\label{eq:Tj}
\end{equation}
According to the definition of $T_j$, the quantities $T_1$ through
$T_7$ are evaluated individually, and the element with the largest
absolute value is selected.  The corresponding index $j^{*}$
identifies the qumode on which the error occurs as
\begin{equation}
j^{*} = \arg \max_{j} \lvert T_j \rvert.
\label{eq:jstar}
\end{equation}

After the location of the error is identified, the magnitude of the
displacement error $\hat{d}_{j^{*}}$ can be estimated as follows. By
computing $T_j$ and $\hat{d}_{j^{*}}$ in parallel, error localization
and magnitude estimation can be efficiently performed.
\begin{equation}
\hat{d}_{j^{*}} = 
\frac{\mathbf{m}_{j^{*}}^{\top} \mathbf{s}_x}
{\lVert \mathbf{m}_{j^{*}} \rVert^{2}}.
\label{eq:d_est}
\end{equation}

\subsection{Logical Operation}
To achieve universal quantum computation, it is necessary to implement
fundamental quantum operations within the encoded Hilbert
space~\cite{Walshe2020,Hu2019}.  For CV systems, the basic quantum
gates consist of the following Gaussian
operations~\cite{Weedbrook2012}: the displacement gate, the rotation
gate, the single-mode squeezing gate, and the beam splitter.  Other
commonly used Gaussian operations, such as the two-mode squeezing gate
and the SUM gate, can be constructed from combinations of beam
splitters and single-mode squeezers, and are therefore not regarded as
fundamental gates~\cite{Hao2014}.

Gaussian operations preserve the Gaussian nature of states. However, 
universal quantum computation requires the inclusion of at least one
non-Gaussian operation.
A typical non-Gaussian operation is the cubic-phase
gate~\cite{Kalajdzievski2021}. In many practical architectures, the cubic-phase gate is not implemented as a native operation, but is instead typically realized via magic-state injection. In this approach, a specially
prepared non-Gaussian resource state (the so-called magic state) is
combined with Gaussian operations, measurement, and feed-forward to
indirectly implement the cubic-phase
transformation~\cite{Walschaers2021}.  Consequently, the entire
framework of universal quantum computation relies fundamentally on
four Gaussian primitive gates.

\begin{figure}[t]
\centering
\includegraphics[width=\linewidth]{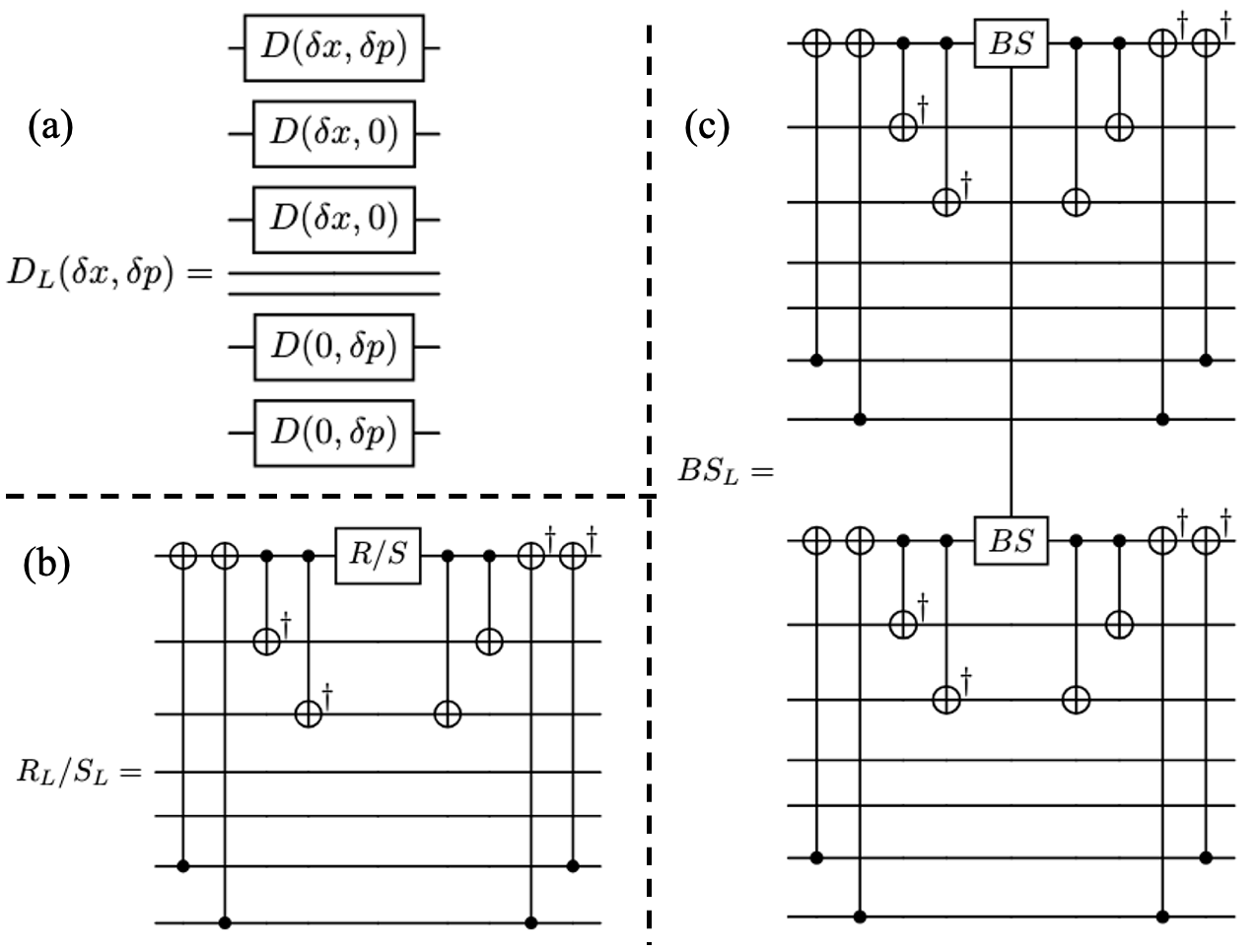}
\caption{\label{fig:logical_op}%
  Logical-operation circuits for (a) the displacement gate, (b) the
  rotation and squeezing gates, and (c) the beam splitter.}
\end{figure}

Figure~\ref{fig:logical_op} shows the implementation circuits of the
four fundamental logical Gaussian gates.  In the DV Steane code, all
logical Pauli operations are transversal and therefore inherently
fault tolerant.  In contrast, for the analog Steane code, the Hilbert
space of each qumode lacks the periodic structure inherent to qubits,
making it difficult to design transversal logical gates and thus to
achieve fault-tolerant quantum computation.  As illustrated in
Fig.~\ref{fig:logical_op}, only the logical displacement operation is
fault tolerant, owing to its transversality, while other logical gates
rely on entangling operations and are therefore non-fault tolerant
(non-transversal).  In the DV Steane code, all logical Clifford
operations can be implemented transversally.

This arises from the CSS structure defined over the finite field
$\mathbb{F}_2$, where the encoding matrix is orthogonal modulo~2.
Local Clifford operations acting independently on each physical qubit
induce the same logical transformations and preserve the stabilizer
structure.  In contrast, in the analog Steane code, logical Clifford
operations correspond to Gaussian unitaries represented by real
symplectic transformations in phase space.  Except for the
displacement gate, Gaussian operations, such as rotation, squeezing,
and beam splitter, mix canonical variables across modes, leading to
non-block-diagonal symplectic forms that destroy the tensor-product
structure of the encoded subspace.  Consequently, these operations
cannot be applied independently to each mode and are nontransversal.
Essentially, DV logical operations rely on discrete modular algebra
that preserves mode independence, whereas CV logical operations are
governed by continuous linear symplectic transformations that
inherently involve mode coupling.  As a result, transversal Clifford
gates exist in the DV Steane code, while only the displacement gate
retains transversality in the analog Steane code. It is important to
emphasize that no claim is made that, in CV systems, transversal
realizations are fundamentally restricted to the logical displacement
gate alone. The statement above pertains specifically to the present
construction and the current state of understanding. It is possible
that future developments may uncover transversal implementations for
additional logical gates within CV encoding frameworks.

Designing fault-tolerant quantum error-correcting codes that can
encode continuous logical information remains a significant challenge.
A promising direction may involve hybrid CV-DV encodings that exploit
the intrinsic periodicity of discrete-variable systems to overcome
this limitation. Although these logical Gaussian gates are not
transversal, they remain implementable within the encoded Hilbert
space and can be interleaved with syndrome extraction. The present
framework therefore supports universal logical operations, but
operates in the regime of strong error suppression rather than
threshold-based fault tolerance, since residual Gaussian noise cannot
be reduced arbitrarily in continuous-variable systems.

\section{Gaussian Error Suppression and Concatenation to Exploit Code Duality}
\label{sec:gaussian_concatenation}

To circumvent the constraint imposed by the no-go
theorem~\cite{Niset2009}, we introduce GKP states as non-Gaussian
resources to suppress Gaussian displacement errors acting on all
qumodes. For residual error beyond the suppression capability of the
GKP layer, a concatenated analog Steane code is employed to further
correct the remaining errors. In combination, these codes create a
functional duality to correct arbitrary displacement errors, i.e., by
providing additional functionality. In contrast, prior work on
concatenation of quantum error correction codes aimed at lower logical
error rates applying functionally equivalent techniques twice.

\subsection{Gaussian Error Suppression Circuit}

\begin{figure}[t]
\centering
\includegraphics[width=\linewidth]{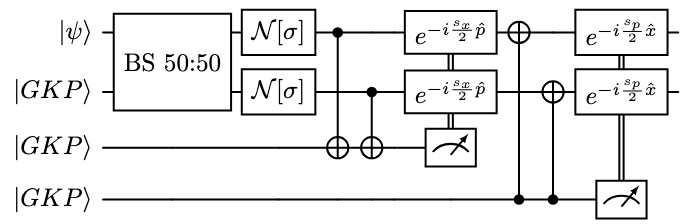}
\caption{\label{fig:gaussian_circuit}%
Gaussian error suppression circuit.}
\end{figure}

A 50:50 beam splitter is employed to entangle a data qumode with a GKP
ancilla, as illustrated in Fig.~\ref{fig:gaussian_circuit}. Gaussian
displacement noise with a standard deviation of $\sigma$ is injected
into both the data and GKP qumodes, as indicated by the
$\mathcal{N}[\sigma]$ symbols in the figure. For syndrome extraction,
position and momentum eigenstates are respectively introduced to read
out the syndrome outcomes on the position and momentum quadratures.

The Gaussian displacement errors acting on the two qumodes are denoted as 
\begin{equation}
[\epsilon_{x,\mathrm{data}},\, \epsilon_{p,\mathrm{data}},\, \epsilon_{x,\mathrm{GKP}},\, \epsilon_{p,\mathrm{GKP}}].
\end{equation}
All displacement errors are assumed to follow independent Gaussian
distributions, i.e.,
\begin{equation}
\epsilon \sim \mathcal{N}(0,\, \sigma^2),
\end{equation}
where $\sigma$ characterizes the standard deviation of the Gaussian
noise in each quadrature.  For an ideal GKP state, the spacing between
adjacent wave packets is $2\sqrt{\pi}$. The corresponding expressions
for the two syndromes in Fig.~\ref{fig:gaussian_circuit} are therefore
given by
\begin{align}
s_x &= R_{2\sqrt{\pi}}\!\left(\epsilon_{x,\mathrm{data}} + \epsilon_{x,\mathrm{GKP}}\right), \label{eq:sx} \\
s_p &= R_{2\sqrt{\pi}}\!\left(-\epsilon_{p,\mathrm{data}} - \epsilon_{p,\mathrm{GKP}}\right), \label{eq:sp}
\end{align}
where $R_{2\sqrt{\pi}}(\cdot)$ denotes the modulo operation that maps
a real variable into the interval $[-\sqrt{\pi},\, \sqrt{\pi})$ with a
period of $2\sqrt{\pi}$.

By subsequently applying displacement gates on the data and
GKP qumodes, Gaussian errors can be effectively suppressed. In
Sec.~IV, we demonstrate that this suppression reduces the variance of
the original Gaussian noise by a factor of two.

For the logical operations of this circuit, since the encoding
operation consists only of a 50:50 beam splitter, all corresponding
logical operations on the data mode can be expressed as
\begin{equation}
\hat{U}_L = \hat{B}^\dagger_{50:50} \, \hat{U} \, \hat{B}_{50:50},
\end{equation}
where $\hat{B}_{50:50}$ denotes the 50:50 beam splitter operation and
$\hat{U}$ represents the corresponding physical-level operation.

\subsection{Code Concatenation for Duality in Displacement}

When the magnitude of a Gaussian displacement exceeds half of the GKP
lattice spacing, the modulo operation maps the result into an
incorrect interval, leading to a lattice-crossing error. Although such
events occur with a relatively low probability, their impact is
typically catastrophic, as they further amplify the magnitude of the
underlying displacement error. Consequently, these errors lie beyond
the correction capability of the circuit shown in
Fig.~\ref{fig:gaussian_circuit}. Our contribution here is to
complement GPK via a concatenation construction with an outer analog
Steane code, which can correct this type of large displacement error,
thereby realizing duality in displacement functionality.

\begin{figure}[t]
\centering
\includegraphics[width=0.55\linewidth]{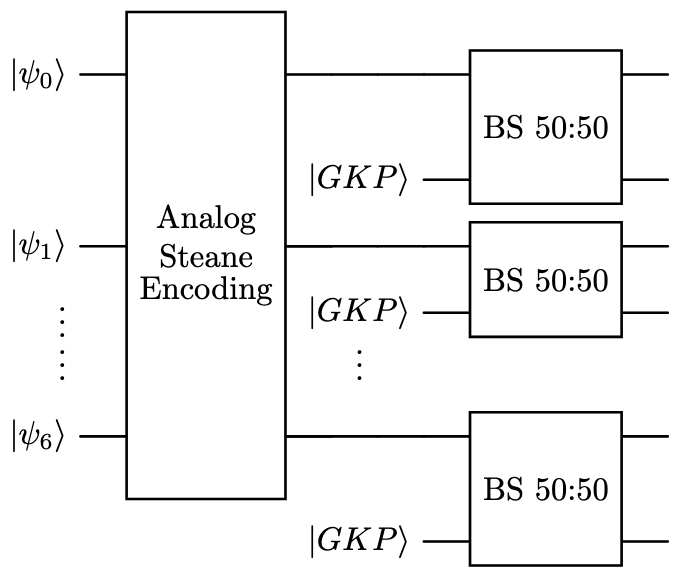}
\caption{Schematic diagram of the concatenated circuit combining the
  analog Steane encoding and the Gaussian error-suppression circuit.}
\label{fig:concatenating}
\end{figure}

Specifically, the circuit is constructed by concatenating the analog
Steane code shown in Fig.~\ref{fig:Steane}(b) with the Gaussian
error-suppression circuit illustrated in
Fig.~\ref{fig:gaussian_circuit}. A schematic diagram of the resulting
concatenated circuit is shown in Fig.~\ref{fig:concatenating}. For
each qumode within the encoding space of Fig.~\ref{fig:Steane}(b), a
50:50 beam splitter is used to entangle it with a GKP state, thereby
forming the concatenated encoding circuit. This circuit naturally
consists of two layers: the inner Gaussian error-suppression layer and
the outer analog Steane layer. During the syndrome-extraction stage,
the inner layer first extracts syndromes to suppress Gaussian errors,
while the outer analog Steane code detects and corrects other
displacement errors.

For logical operations, since both the analog Steane code and the Gaussian
error-suppression circuit have their respective logical operations
well defined, the overall logical operation can be obtained by
combining the two accordingly.

Previous concatenation schemes for bosonic codes, such as the
GKP--repetition construction of Li and Su~\cite{Li2024NoisyAncilla},
convert biased Gaussian displacement noise into a discrete biased
Pauli-$\bar{X}$ channel via a round of GKP error correction. They then
apply a qubit-level repetition code with projective syndrome
extraction that classifies measurement outcomes into Pauli-error zones
and no-Pauli-error zones (i.e., by determining whether the residual
displacement falls into phase-space regions associated with a logical
Pauli flip or not) followed by digital majority voting. In contrast,
our architecture operates at a fundamentally different level. We
utilize the GKP layer solely as a Gaussian error-suppression mechanism
that continuously reduces displacement variance on each mode, without
collapsing errors into discrete Pauli events. Furthermore, the outer
analog Steane code directly encodes continuous-variable logical
information across multiple modes. Syndrome information in our scheme
is obtained entirely through Gaussian circuits and homodyne
detection. It is then processed as a full analog syndrome vector,
enabling simultaneous inference of both the location and the magnitude
of displacement errors, rather than only detecting whether a logical
Pauli flip has occurred. This analog, CV-level decoding --- combined
with the dual-displacement structure of the code --- allows our
concatenated dual displacement code to jointly suppress small Gaussian
fluctuations and rare lattice-crossing shifts within a single CV
framework and using only one class of non-Gaussian resource (GKP
ancillas), thereby going beyond DV-oriented GKP--repetition
concatenations in both syndrome extraction strategy and operational
scope.

\section{Performance Evaluation}
\label{sec:performance}
This Section is organized as follows.    
Subsection~\ref{subsec:error_model} introduces the error model.  
Subsection~\ref{subsec:ideal_analysis} analyzes the performance of the concatenated code in the idealized setting,  
while Subsection~\ref{subsec:real_analysis} examines its performance under realistic conditions.  
Subsection~\ref{subsec:exp_feasibility} provides a detailed assessment of the experimental feasibility of implementing the concatenated code.  
Subsection~\ref{subsec:simulation_results} presents Monte--Carlo simulations that evaluate the code performance.  
Finally, Subsection~\ref{subsec:qubit_comparison} compares the proposed oscillator-based approach with schemes that encode an oscillator using qubits.

\subsection{Error Model}
\label{subsec:error_model}

Our concatenated scheme is inherently compatible with superconducting
cavity-QED and trapped-ion platforms that feature long-lived bosonic
modes, high-fidelity Gaussian operations, and access to non-Gaussian
GKP resources.  In the weak-noise regime, photon-loss,
thermal-excitation, and dephasing channels can all be effectively
modeled as Gaussian displacement noise in phase space, since their
cumulative effect corresponds to random quadrature displacements
characterized by Gaussian statistics~\cite{Wu2021}. Therefore, one of
the primary noise sources considered in this work is Gaussian
displacement error. Gaussian displacement error can be described as a
statistical mixture of phase-space displacement operators whose
amplitudes follow a Gaussian probability
distribution~\cite{Ralph2011}. The Gaussian displacement noise channel
acts on any density matrix $\rho$ as
\begin{equation}
\mathcal{E}_{\mathrm{Gauss}}(\rho) = \int P(\delta)\, D(\delta)\, \rho\, D^{\dagger}(\delta)\, d\delta,
\end{equation}
where $P(\delta)$ denotes a Gaussian distribution over displacements
$\delta$.

According to the no-go theorem, Gaussian displacement errors cannot be
effectively suppressed using only Gaussian elements. Therefore, a
Gaussian error-suppression circuit was designed in Sec.~III. However,
this circuit can suppress only those Gaussian errors whose magnitude
does not exceed half of the GKP lattice spacing. Otherwise, a
lattice-crossing error occurs.

In addition to Gaussian displacement noise, the proposed concatenated
scheme can also correct occasional large-amplitude displacement
errors, which we refer to as abrupt displacement errors. These rare
but significant errors often stem from control pulse miscalibration,
sudden flux or charge jumps, and other non-Gaussian noise events that
intermittently disturb the oscillator
dynamics~\cite{Jiang2018,Li2025}.
Unlike Gaussian noise, these events introduce abrupt, localized shifts
in the quadrature amplitudes that can exceed the typical Gaussian
error variance and lead to decoding failure if uncorrected.

\subsection{Analysis under Ideal Conditions}
\label{subsec:ideal_analysis}

The ideal condition refers to the case where all eigenstates and GKP
states are assumed to be infinitely squeezed. The analysis under this
assumption represents the theoretical limit that the code can achieve.

\subsubsection{Gaussian Error Suppression}
After the feedforward operation shown in
Fig.~\ref{fig:gaussian_circuit}, the residual displacement errors on
the data and GKP qumodes are given by
\begin{align}
\xi_{x,\mathrm{data}}^{(\mathrm{out})} &= \epsilon_{x,\mathrm{data}} - \tfrac{1}{2} R_{2\sqrt{\pi}}\!\left(\epsilon_{x,\mathrm{data}} + \epsilon_{x,\mathrm{GKP}}\right), \label{eq:xi_x_data_out} \\
\xi_{p,\mathrm{data}}^{(\mathrm{out})} &= \epsilon_{p,\mathrm{data}} - \tfrac{1}{2} R_{2\sqrt{\pi}}\!\left(\epsilon_{p,\mathrm{data}} + \epsilon_{p,\mathrm{GKP}}\right), \\
\xi_{x,\mathrm{GKP}}^{(\mathrm{out})}  &= \epsilon_{x,\mathrm{GKP}} - \tfrac{1}{2} R_{2\sqrt{\pi}}\!\left(\epsilon_{x,\mathrm{data}} + \epsilon_{x,\mathrm{GKP}}\right), \\
\xi_{p,\mathrm{GKP}}^{(\mathrm{out})}  &= \epsilon_{p,\mathrm{GKP}} - \tfrac{1}{2} R_{2\sqrt{\pi}}\!\left(\epsilon_{p,\mathrm{data}} + \epsilon_{p,\mathrm{GKP}}\right).
\end{align}

It can be seen that the expressions for each quadrature on the two
qumodes are completely identical, forming a fully symmetric
error-correction structure. Taking Eq.~(\ref{eq:xi_x_data_out}) as an
example, the probability density function of the $x$~quadrature on the
data qumode is given by
\begin{equation}
\resizebox{\linewidth}{!}{$
\begin{aligned}
X\!\left(\xi_{x,\mathrm{data}}^{(\mathrm{out})}\right)
&= \frac{1}{2\sqrt{\pi}\,\sigma}
\sum_{m\in\mathbb{Z}}
\exp\!\left[-\frac{\big(m\sqrt{\pi}-\xi_{x,\mathrm{data}}^{(\mathrm{out})}\big)^{2}}{\sigma^{2}}\right]\\
&\quad\times
\Bigg[
\operatorname{erf}\!\left(\frac{m\sqrt{\pi}+\frac{\sqrt{\pi}}{2}}{\sigma}\right)
-\operatorname{erf}\!\left(\frac{m\sqrt{\pi}-\frac{\sqrt{\pi}}{2}}{\sigma}\right)
\Bigg].
\end{aligned}
$}
\label{eq:X_chain}
\end{equation}
where $\operatorname{erf}(x) = \frac{2}{\sqrt{\pi}}\!\int_0^x e^{-t^2}\,dt$ is the error function.

Since $X\!\left(\xi_{x,\mathrm{data}}^{(\mathrm{out})}\right)$ is an
even function, the mean value of the corresponding probability density
function is
$\mathbb{E}\!\left[\xi_{x,\mathrm{data}}^{(\mathrm{out})}\right] =
\int \xi_{x,\mathrm{data}}^{(\mathrm{out})}
X\!\left(\xi_{x,\mathrm{data}}^{(\mathrm{out})}\right)
d\xi_{x,\mathrm{data}}^{(\mathrm{out})} = 0.$

The variance of the corresponding probability density function is given by  
\begin{equation}
\resizebox{\linewidth}{!}{$
\begin{aligned}
\mathrm{Var}\!\left(\xi_{x,\mathrm{data}}^{(\mathrm{out})}\right)
&= \frac{1}{2}
\sum_{m\in\mathbb{Z}}
\!\left(m^{2}\pi+\frac{\sigma^{2}}{2}\right) \\
&\quad\times
\Bigg[
\operatorname{erf}\!\left(\frac{m\sqrt{\pi}+\frac{\sqrt{\pi}}{2}}{\sigma}\right)
-
\operatorname{erf}\!\left(\frac{m\sqrt{\pi}-\frac{\sqrt{\pi}}{2}}{\sigma}\right)
\Bigg].
\end{aligned}
$}
\label{eq:var_final}
\end{equation}
A detailed derivation is provided in Appendix~\ref{app:derivation}, spanning Eqs.~(\ref{eq:X_chain_detailed})-(\ref{eq:var_final_detail}).

When the standard deviation of the Gaussian displacement noise is
small ($\sigma \!\ll\! 1$), the $m=0$ term dominates the summation,
and contributions from $m \neq 0$ can be neglected. Therefore, the
variance of the residual Gaussian noise can be expressed as
\begin{equation}
\resizebox{\linewidth}{!}{$
\sigma_{\mathrm{res}}^{2}
= \sigma_{\mathrm{res},x}^{2}
= \sigma_{\mathrm{res},p}^{2}
= \mathrm{Var}\!\left(\xi_{x,\mathrm{data}}^{(\mathrm{out})}\right)
\approx \tfrac{1}{2}\!\left(0+\tfrac{\sigma^{2}}{2}\right)\!\times 2
= \tfrac{\sigma^{2}}{2}.
$}
\label{eq:sigma_res}
\end{equation}

This result indicates that the Gaussian noise suppression circuit
reduces the variance to half of its original value.  However, it
should be noted that the residual Gaussian noise can accumulate over
successive QEC cycles, which may eventually lead to failure in
long-duration quantum computation tasks due to the cumulative effect
of noise.

\subsubsection{Concatenate Code}
The lattice structure of the GKP states enables the suppression of
small displacement errors by correcting shifts within each unit cell.
However, when the displacement magnitude exceeds half of the lattice
spacing, a lattice-crossing event occurs. For the proposed
concatenated code, the outer analog Steane code is responsible for
correcting the lattice-crossing and abrupt errors.  The composition of
the analog Steane code's syndrome can be expressed as follows.  Taking the
position-quadrature syndrome as an example,
\begin{equation}
\mathbf{s}_{x} = \mathbf{M}_{x}\,\boldsymbol{\epsilon}_{\mathrm{res},x} + d\,\mathbf{m}_{j}.
\label{eq:cvsteane_sx}
\end{equation}
Here, $d$ denotes the magnitude of the lattice-crossing error, and
$\boldsymbol{\epsilon}$ represents the residual Gaussian noise, which
follows a normal distribution
$\mathcal{N}(0, \sigma_{\mathrm{res}}^{2})$.

Equations~\eqref{eq:Tj} and~\eqref{eq:d_est} show the procedures for
error localization and magnitude estimation without considering the
residual Gaussian noise.  When such residual noise is taken into
account, the data should be whitened to improve the accuracy of both
error localization and magnitude estimation~\cite{Kessy2018}.  Let the
covariance matrix of the position-quadrature syndrome be denoted as
$\boldsymbol{\Sigma}_{s_{x}}$, and define the whitening matrix as
$\mathbf{W} = \boldsymbol{\Sigma}_{s_{x}}^{-1/2}$, yielding

\begin{equation}
T_{j} = 
\frac{
\mathbf{m}_{j}^{\mathsf{T}} \boldsymbol{\Sigma}_{s_{x}}^{-1} \mathbf{s}_{x}
}{
\sqrt{ \mathbf{m}_{j}^{\mathsf{T}} \boldsymbol{\Sigma}_{s_{x}}^{-1} \mathbf{m}_{j} }
},
\label{eq:Tj_W}
\end{equation}

\begin{equation}
\hat{d}_{j^{\ast}} =
\frac{
\mathbf{m}_{j^{\ast}}^{\mathsf{T}} \boldsymbol{\Sigma}_{s_{x}}^{-1} \mathbf{s}_{x}
}{
\mathbf{m}_{j^{\ast}}^{\mathsf{T}} \boldsymbol{\Sigma}_{s_{x}}^{-1} \mathbf{m}_{j^{\ast}}
}.
\label{eq:d_est_W}
\end{equation}

For Eq.~\eqref{eq:d_est}, the expectation and variance of its
estimator are given by
\begin{equation}
\mathbb{E}\!\left[\hat{d}_{j^{\ast}}\right] = d, \qquad
\mathrm{Var}\!\left(\hat{d}_{j^{\ast}}\right) =
\frac{1}{
\mathbf{m}_{j^{\ast}}^{\mathsf{T}} \boldsymbol{\Sigma}_{s_{x}}^{-1} \mathbf{m}_{j^{\ast}}
}.
\end{equation}

For qumodes within the analog Steane code block that do not experience
lattice-crossing or abrupt errors, the position or momentum variance
after one round of correction remains equal to the residual variance
$\sigma_{\mathrm{res}}^{2}$ from the inner Gaussian-suppression
circuit.  For qumodes that experience a lattice-crossing error, the
residual variance of their position or momentum quadrature is given by
\begin{equation}
\begin{aligned}
\mathrm{Var}\!\left(\mathrm{qumode}_{j^{\ast}}\right)
&= \sigma_{\mathrm{res}}^{2} + \mathrm{Var}\!\left(\hat{d}_{j^{\ast}}\right) - 2\,\mathrm{Cov}\!\left(\epsilon_{j^{\ast}}, \hat{d}_{j^{\ast}}\right) \\[2pt]
&= \sigma_{\mathrm{res}}^{2} + \mathrm{Var}\!\left(\hat{d}_{j^{\ast}}\right) - 2\sigma_{\mathrm{res}}^{2} \\[2pt]
&= \mathrm{Var}\!\left(\hat{d}_{j^{\ast}}\right) - \sigma_{\mathrm{res}}^{2}.
\end{aligned}
\end{equation}

For the position quadrature, when lattice-crossing or abrupt errors
are sequentially considered on each qumode, the value of
$\mathrm{Var}\!\left(\hat{d}_{j^{\ast}}\right) -
2\sigma_{\mathrm{res}}^{2}$ is summarized in
Table~\ref{tab:variance_change}. It can be observed that when
lattice-crossing or abrupt errors occur on qumodes~4, 6, and~7, the
residual variances on these qumodes after error correction are lower
than the residual Gaussian noise.  This observation indicates that the
analog Steane code is not a perfectly symmetric encoding structure.
\begin{table}[htbp]
\centering
\caption{Variance changes under lattice-crossing errors.}
\begin{tabular}{c | c}
\hline
$j^{\ast}$ &
$\mathrm{Var}\!\left(\hat{d}_{j^{\ast}}\right) - 2\sigma_{\mathrm{res}}^{2}$ \\ 
\hline
1 & $\sigma_{\mathrm{res}}^{2}$ \\
2 & $\sigma_{\mathrm{res}}^{2}$ \\
3 & $\sigma_{\mathrm{res}}^{2}$ \\
4 & $-\tfrac{2}{7}\sigma_{\mathrm{res}}^{2}$ \\
5 & $2\sigma_{\mathrm{res}}^{2}$ \\
6 & $-\tfrac{2}{7}\sigma_{\mathrm{res}}^{2}$ \\
7 & $-\tfrac{2}{7}\sigma_{\mathrm{res}}^{2}$ \\
\hline
\end{tabular}
\label{tab:variance_change}
\end{table}

Another case must be considered: When the magnitude of a
lattice-crossing or abrupt error becomes comparable to the residual
Gaussian noise, the whitened error localization defined in
Eq.~\eqref{eq:Tj_W} may lead to a miscorrection event.  Miscorrection
refers to the case where the error identified through the syndrome
does not coincide with the actual physical error that occurred.
Assume that the lattice-crossing or abrupt error actually occurs on
qumode~$j$ but is misidentified as occurring on qumode~$k$.  Let the
whitened position syndrome and corresponding pattern be
$\tilde{\mathbf{s}}_{x} = \mathbf{W}\mathbf{s}_{x}$ and
$\tilde{\boldsymbol{\mu}}_{j} = d\,\mathbf{W}\mathbf{m}_{j}$,
respectively.  If qumode~$j$ is correctly identified, then
\begin{equation}
\resizebox{\linewidth}{!}{$
\|\tilde{\mathbf{s}}_{x} - \tilde{\boldsymbol{\mu}}_{j}\|^{2}
\leq
\|\tilde{\mathbf{s}}_{x} - \tilde{\boldsymbol{\mu}}_{k}\|^{2}
\;\Longleftrightarrow\;
\left(\tilde{\boldsymbol{\mu}}_{j} - \tilde{\boldsymbol{\mu}}_{k}\right)^{\mathsf{T}}
\!\left(\tilde{\mathbf{s}}_{x} - \frac{\tilde{\boldsymbol{\mu}}_{j} + \tilde{\boldsymbol{\mu}}_{k}}{2}\right)
\ge 0.
$}
\end{equation}

Define
$g_{jk} =
\left(\tilde{\boldsymbol{\mu}}_{j} - \tilde{\boldsymbol{\mu}}_{k}\right)^{\mathsf{T}}
\!\left(\tilde{\mathbf{s}}_{x} - \frac{\tilde{\boldsymbol{\mu}}_{j} + \tilde{\boldsymbol{\mu}}_{k}}{2}\right)$,
which follows a Gaussian distribution:
\begin{equation}
\begin{aligned}
g_{jk} &\sim 
\mathcal{N}\!\left(
\tfrac{1}{2}\|\boldsymbol{\delta}_{jk}\|^{2},
\|\boldsymbol{\delta}_{jk}\|^{2}
\right), \\
\boldsymbol{\delta}_{jk}
&:= \tilde{\boldsymbol{\mu}}_{j} - \tilde{\boldsymbol{\mu}}_{k}
= d\,\mathbf{W}\left(\mathbf{m}_{j} - \mathbf{m}_{k}\right).
\end{aligned}
\end{equation}

When qumode~$j$ is misidentitied as qumode~$k$, the probability of this
event is given by
\begin{equation}
\Pr(j \!\rightarrow\! k)
= \Pr(g_{jk} < 0)
= Q\!\left(\tfrac{1}{2}\|\boldsymbol{\delta}_{jk}\|\right),
\end{equation}
where
$Q(x) = \tfrac{1}{\sqrt{2\pi}}\!\int_{x}^{\infty} e^{-t^{2}/2}\,dt$ is
the right tail of the standard normal distribution. Expanding the
expression of $\boldsymbol{\delta}_{jk}$ yields
\begin{equation}
\|\boldsymbol{\delta}_{jk}\|^{2}
= d^{2}(\mathbf{m}_{j} - \mathbf{m}_{k})^{\mathsf{T}}
\boldsymbol{\Sigma}_{s_{x}}^{-1}(\mathbf{m}_{j} - \mathbf{m}_{k})
= d^{2}\,\Delta_{jk}^{2}.
\end{equation}
\begin{equation}
\Pr(j \!\rightarrow\! k)
= Q\!\left(\tfrac{d}{2}\sqrt{\Delta_{jk}^{2}}\right).
\label{eq:mislocation_prob}
\end{equation}

According to Eq.~\eqref{eq:mislocation_prob}, the probability of a
miscorrection event where qumode~$j$ is identified as all $k \neq j$
satisfies
\begin{equation}
P_{\mathrm{miscorr}}^{(j)} 
\leq \sum_{k \neq j} 
Q\!\left(\tfrac{d}{2}\sqrt{\Delta_{jk}^{2}}\right).
\label{eq:mislocation_union}
\end{equation}

\begin{figure*}[t]
    \centering
    \includegraphics[width=0.8\textwidth]{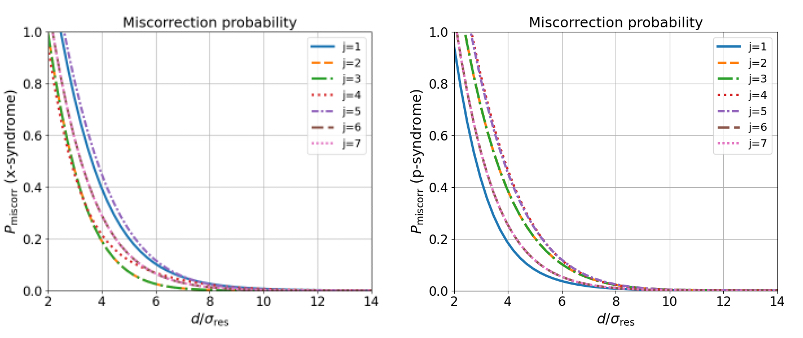}
    \caption{Miscorrection probability $P_{\mathrm{miscorr}}^{(j)}$ as
      a function of the normalized ratio
      $d/\sigma_{\mathrm{res}}$. The index $j=1,\dots,7$ labels the
      qumode position within the 7-mode analog Steane encoding block
      (ordered from top to bottom in the encoding circuit). A
      miscorrection event refers to the situation in which a
      displacement error occurring on the $j$-th qumode is incorrectly
      localized due to residual Gaussian displacement noise. In the
      calculation, a displacement error of magnitude $d$ is applied to
      a given qumode $j$, while each mode carries residual Gaussian
      displacement noise with standard deviation
      $\sigma_{\mathrm{res}}$. The analog Steane code attempts to
      identify the error location using its syndrome
      information. Gaussian errors may distort the inferred syndrome
      values, leading to a nonzero probability of incorrect
      localization. Each curve shows this probability for a given
      $j$. The left panel corresponds to the position quadrature
      ($x$), and the right panel corresponds to the momentum
      quadrature ($p$). The miscorrection probability decreases as
      $d/\sigma_{\mathrm{res}}$ increases, since larger displacement
      magnitudes are less affected by the same level of residual
      Gaussian noise~\cite{guo2026cvqec}.}
    \label{fig:Pmisloc}
\end{figure*}

As shown in Fig.~\ref{fig:Pmisloc}, the miscorrection probability
$P_{\mathrm{miscorr}}^{(j)}$ in both the position and momentum
quadratures depends on the ratio between the displacement error
magnitude $d$ and the residual Gaussian noise standard deviation
$\sigma_{\mathrm{res}}$.  As $d/\sigma_{\mathrm{res}}$ increases,
$P_{\mathrm{miscorr}}^{(j)}$ decreases rapidly.  When
$d/\sigma_{\mathrm{res}} \approx 10$, the value of
$P_{\mathrm{miscorr}}^{(j)}$ approaches zero, indicating that
miscorrection events are essentially eliminated.

\subsection{Analysis under Real Conditions}
\label{subsec:real_analysis}

In realistic conditions, both eigenstates and GKP states are finitely
squeezed, which introduces additional uncertainty into the proposed
concatenated code.  For the inner Gaussian-error-suppression circuit,
a finitely squeezed GKP state can be regarded as an ideal GKP state
superimposed with an additional Gaussian noise whose variance depends
on the squeezing strength.  Similarly, the position or momentum
eigenstates become finitely squeezed vacuum states.

Let the squeezing parameter be denoted by $r$.  The variance along the
squeezed quadrature can then be expressed as $\tfrac{1}{2}e^{-2r}$,
which also corresponds to the variance of each peak in a finitely
squeezed GKP state. Since the global Gaussian envelope width of GKP
state does not influence syndrome extraction, it is not included in
the analysis throughout this work. Similar to
Eqs.~(\ref{eq:X_chain})-(\ref{eq:sigma_res}), under finite squeezing,
the residual Gaussian noise variance on the data qumodes of the
Gaussian-error-suppression circuit is given by
\begin{equation}
\sigma_{\mathrm{res}}^{2}
= \tfrac{1}{2}\sigma^{2}
+ \tfrac{1}{8}e^{-2r}
+ \mathcal{O}\!\left(R_{2\sqrt{\pi}}(\cdot)\right),
\label{eq:residual_variance_real}
\end{equation}
where the term $\mathcal{O}\!\left(R_{2\sqrt{\pi}}(\cdot)\right)$
represents the additional error contribution caused by lattice
crossing, which becomes negligible when $\sigma$ is small.  The second
term in Eq.~(38), $\frac{1}{8}e^{-2r}$, explicitly quantifies the
residual noise contribution arising from finite GKP squeezing. As the
squeezing parameter $r$ increases, this term decays exponentially,
indicating that the achievable Gaussian-noise suppression improves
continuously with GKP quality. Conversely, for smaller $r$, the
residual variance increases smoothly, leading to a gradual reduction
of suppression gain rather than a sudden failure of the correction
mechanism.
Therefore, from Eq.~\eqref{eq:residual_variance_real}, it can be seen
that an error-suppression gain can be achieved when
$r > -\ln(2\sigma)$.

For the outer analog Steane code, taking the position quadrature as an
example, the expression for the syndrome is modified from
Eq.~\eqref{eq:cvsteane_sx} to
\begin{equation}
\mathbf{s}_{x} = 
\mathbf{M}_{x}\boldsymbol{\epsilon}_{\mathrm{res},x}
+ d\,\mathbf{m}_{j}
+ \mathbf{A}_{x}\mathbf{n}_{x}.
\label{eq:new_sx}
\end{equation}
where 
$\mathbf{n}_{x} = [\hat{x}_{2}, \hat{x}_{3}, \hat{x}_{4}, \hat{x}_{\mathrm{anc}1}, \hat{x}_{\mathrm{anc}2}, \hat{x}_{\mathrm{anc}3}]^{\mathsf{T}}$,  
each position operator has a variance of $\tfrac{1}{2}e^{-2r}$,  
and $\mathbf{A}_{x}$ is given by  
\begin{equation}
\mathbf{A}_{x} =
\begin{bmatrix}
1 & 0 & 0 & 1 & 0 & 0 \\
0 & 1 & 0 & 0 & 1 & 0 \\
0 & 0 & 1 & 0 & 0 & 1
\end{bmatrix}.
\end{equation}

In the error localization (Eq.~\eqref{eq:Tj_W}) and magnitude
estimation (Eq.~\eqref{eq:d_est_W}), the covariance matrix of
$\mathbf{s}_{x}$ is utilized.  According to Eq.~\eqref{eq:new_sx}, the
updated covariance matrix is given by
\begin{equation}
\boldsymbol{\Sigma}_{s_x}
= \sigma_{\mathrm{res}}^{2}\mathbf{M}_{x}\mathbf{M}_{x}^{\mathsf{T}}
+ \tfrac{1}{2}e^{-2r}\mathbf{A}_{x}\mathbf{A}_{x}^{\mathsf{T}}.
\end{equation}

Through the covariance matrix $\boldsymbol{\Sigma}_{s_x}$, one can
qualitatively analyze how the key parameters vary with the squeezing
parameter $r$.  Equation~\eqref{eq:d_est_W} gives the variance of the
estimated magnitude of the lattice-crossing error, whose derivative is
\begin{equation}
\resizebox{\columnwidth}{!}{$
\begin{aligned}
\frac{d}{dr}\mathrm{Var}(\hat{d}_{j^*})
&= -\frac{1}{\bigl(\mathbf{m}_{j^*}^{T}\boldsymbol{\Sigma}_{s_x}^{-1}\mathbf{m}_{j^*}\bigr)^2}
\,\frac{d}{dr}\!\left(\mathbf{m}_{j^*}^{T}\boldsymbol{\Sigma}_{s_x}^{-1}\mathbf{m}_{j}\right) \\[4pt]
&= -\frac{1}{\bigl(\mathbf{m}_{j^*}^{T}\boldsymbol{\Sigma}_{s_x}^{-1}\mathbf{m}_{j^*}\bigr)^2}
\left( 2e^{-2r}\,\mathbf{m}_{j}^{T}\boldsymbol{\Sigma}_{s_x}^{-2}\mathbf{m}_{j} \right) \;<\; 0 .
\end{aligned}
$}
\label{eq:var_derivative}
\end{equation}
As the squeezing strength increases, the variance of the estimated
error magnitude decreases, indicating that stronger squeezing enhances
the performance of the analog Steane code.  Similarly, one can infer that
the probability of miscorrection also decreases with increasing
squeezing strength, thereby reducing the risk of catastrophic failures
caused by error miscorrection.

\subsection{Experimental Feasibility}
\label{subsec:exp_feasibility}

The concatenated encoding architecture with its functional duality
under consideration employs an analog Steane code as the outer layer to
correct lattice-crossing and abrupt errors, and GKP states as the
inner layer to suppress Gaussian displacement error. Crucially, apart
from the preparation of approximate GKP states, no other non-Gaussian
resources are required.

Squeezed optical modes with moderate levels of squeezing (e.g.,
\(r \sim 10\mbox{-}12\;\mathrm{dB}\)) are already routinely achieved
in modern CV platforms~\cite{Schnabel2017}. In our scheme, the outer
analog Steane code can be implemented entirely with Gaussian operations
(beam splitters, SUM gate, Fourier gate, homodyne detection and
feed-forward) plus the injection of GKP ancilla modes. Since the only
non-Gaussian element is the finite-energy GKP state, the experimental
overhead is significantly reduced compared to schemes relying on large
numbers of non-Gaussian gates.

Recent theoretical works on concatenated GKP-based codes have
established that error suppression becomes feasible once the GKP
squeezing surpasses a threshold on the order of 10-15 dB under
realistic noise assumptions~\cite{Noh2020,Conrad2022}. Given that our
concatenation further leverages the analog Steane code to absorb residual
displacement errors, the required GKP resource quality is relaxed,
making the scheme experimentally accessible in the near term.

We summarize in Table~\ref{tab:platform_noise} the levels of Gaussian
displacement noise variance per single QEC round across different
experimental
platforms~\cite{Huang2024Nature,Epstein2007PRA,Ofek2016Nature,Wang2018PRApplied}.
For some platforms, the reported values are obtained through
subsequent conversions. For instance, the result for the trapped-ion
platform is inferred from the measured motional heating rate.
Table~\ref{tab:platform_noise} presents the corresponding noise
variances and lattice-crossing probabilities for a GKP state with a
squeezing level of \(10~\mathrm{dB}\) across various platforms. The
probability of a lattice-crossing error is evaluated using
$\mathrm{erfc}\!\left(\frac{\sqrt{\pi}}{2\sqrt{2}\sigma}\right)$,
which represents the two-tailed Gaussian probability that the
displacement magnitude exceeds $\sqrt{\pi}/2$.  Among these platforms,
the optomechanical system is significantly affected by Gaussian noise,
which constitutes one of its dominant noise sources.  In contrast,
other platforms exhibit lattice-crossing probabilities well below
$1\%$.  However, as the number of qumodes increases, this probability
grows exponentially, and thus remains non-negligible.

As a concrete example, consider the trapped-ion platform listed in
Table~\ref{tab:platform_noise}, with an effective Gaussian noise
variance of $\sigma^2 \approx 0.03$.  Using the suppression condition
derived from Eq.~(\ref{eq:residual_variance_real}),
\begin{equation}
r > -\ln(2\sigma),
\label{eq:squeezing_requirement}
\end{equation}
we obtain a minimum squeezing requirement of $r > 1.06$,
corresponding to approximately $9.2$\,dB.
This shows that experimentally relevant GKP squeezing levels are already sufficient
to produce observable Gaussian-noise suppression gain in a proof-of-principle
demonstration, while higher squeezing improves performance continuously.

\begin{table}
\caption{Effective Gaussian noise levels and lattice-crossing probabilities
for representative physical platforms.}
\label{tab:platform_noise}
\begin{ruledtabular}
\begin{tabular}{lcc}
Platform &
\makecell[c]{Effective Gaussian\\noise variance} &
\makecell[c]{Lattice-crossing\\probability} \\
\hline
Optical CV~\cite{Wang2018PRApplied}    & $\sim 0.005$ & $0.016\%$ \\
cQED~\cite{Ofek2016Nature}             & $\sim 0.02$  & $0.081\%$ \\
Trapped-ion~\cite{Epstein2007PRA}      & $\sim 0.03$  & $0.17\%$  \\
Optomechanics~\cite{Huang2024Nature}    & $\sim 0.2$   & $7.6\%$    \\
\end{tabular}
\end{ruledtabular}
\end{table}

In a cQED platform, the proposed concatenated code is estimated to
take about \(7~\upmu\mathrm{s}\) for a single round, assuming highly
parallelized gate scheduling.  This estimate is based on
characteristic operation times of \(150~\mathrm{ns}\) for a SUM
gate~\cite{Lu2023NatComm}, \(20~\mathrm{ns}\) for a displacement
gate~\cite{Leghtas2015Science}, and \(1100~\mathrm{ns}\) for a
homodyne measurement~\cite{Bultink2018APL}.  Given the superconducting
resonator lifetimes of \(250\text{--}350~\upmu\mathrm{s}\) achieved on
the Yale cQED platform~\cite{Ofek2016Nature}, multiple QEC rounds can
be executed well within the coherence window, thereby further
enhancing the logical lifetime of the protected information.

Key experimental considerations include:  
Preparation of approximate GKP states with fidelity sufficient to
reduce inner-layer miscorrection probabilities below the outer-layer
decoding threshold is achieved.  Optical losses and finite detection
efficiencies must be kept low enough such that the effective
displacement noise entering the outer Steane layer remains within the
correctable regime.  Feed-forward latency and stability of Gaussian
elements rely on the architecture implementing real-time displacement
corrections based on syndrome measurements, which is well within the
performance of current CV optics platforms.

In summary, the proposed concatenated architecture requires only one
class of non-Gaussian resource (the GKP state) while all other
operations remain Gaussian. With moderate squeezing levels and
high-efficiency Gaussian measurement/detection, the full setup is
within reach of current or near-term continuous-variable quantum
optics experiments.

\subsection{Simulation Results}
\label{subsec:simulation_results}

\begin{figure*}[t]
    \centering
    \includegraphics[width=1\textwidth]{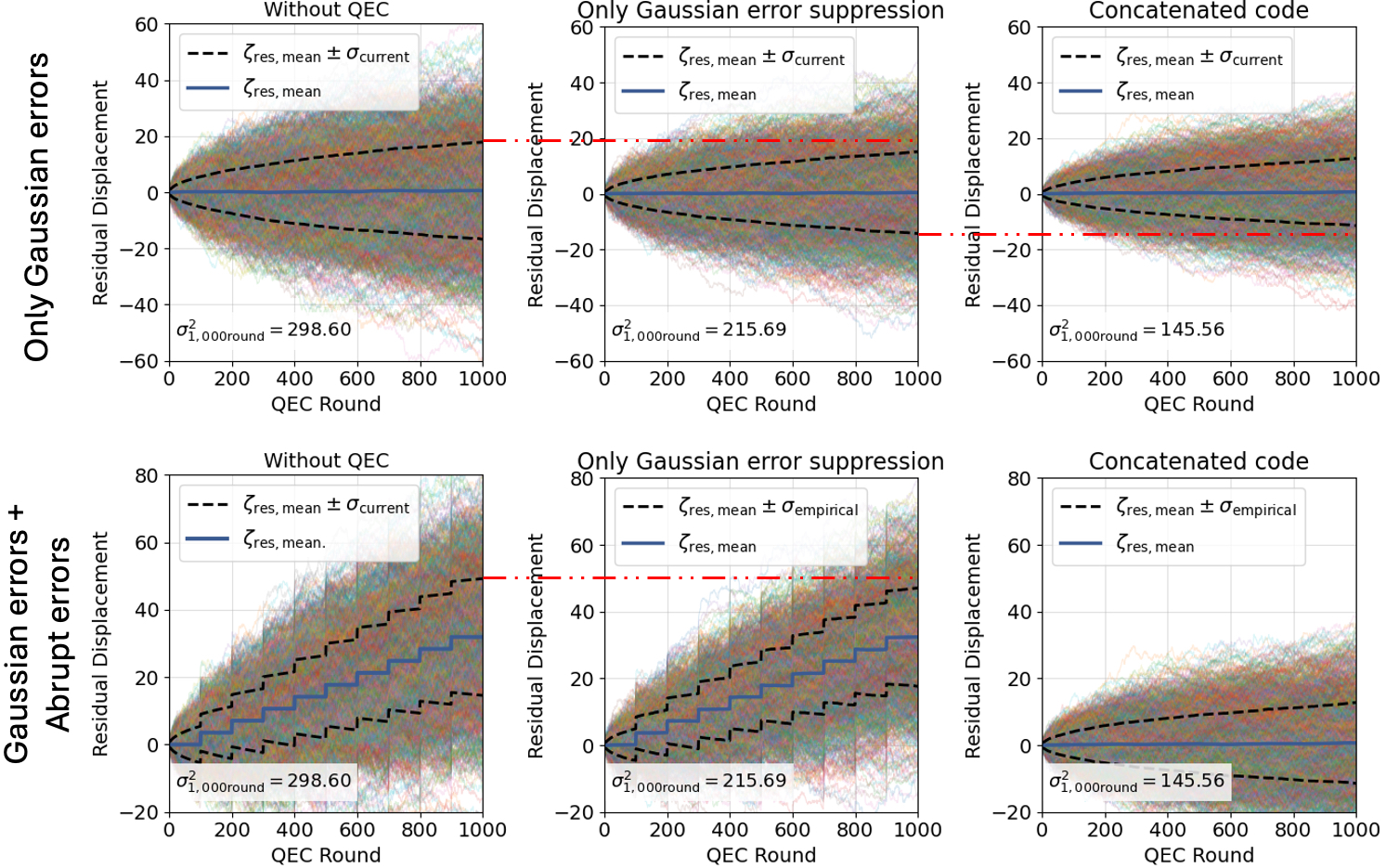}
    \caption{ Monte Carlo simulation of the residual displacement
      $\zeta_{\mathrm{res}}$ over $1{,}000$ QEC rounds.  Each panel
      shows $2{,}000$ individual trajectories (faint curves), the mean
      value $\zeta_{\mathrm{res,mean}}$ (blue solid line), and the
      interval
      $\zeta_{\mathrm{res,mean}} \pm \sigma_{\mathrm{current}}$ (black
      dashed lines), where $\sigma_{\mathrm{current}}$ denotes the
      sample standard deviation in the corresponding round.  The first
      row corresponds to the Gaussian-only noise model, where
      independent displacement noise with variance $0.3$ is applied to
      each quadrature per round.  While the second row includes both
      Gaussian errors and periodic abrupt displacement errors of fixed
      magnitude $+2\sqrt{\pi}$ applied every $100$ QEC rounds.  The
      columns represent, from left to right, the cases of no QEC,
      Gaussian-error suppression only, and the full concatenated code.
      The value shown in the lower-left corner of each panel indicates
      the empirical variance $\sigma_{\mathrm{current}}^{2}$ at the
      end of the $1000$th QEC round~\cite{guo2026cvqec}.  }
    \label{fig:simulation}
\end{figure*}

To illustrate the error-suppression capabilities of different
correction mechanisms under idealized conditions (i.e., infinite squeezing),
Monte Carlo simulations are performed using an error-evolution model.
The goal is to compare how different correction layers reduce displacement
noise, rather than to simulate the full multi-mode quantum circuit.

A full circuit-level simulation of the present bosonic architecture is
not pursued because it would obscure, rather than clarify, the
physical mechanisms that this work aims to isolate. In multi-mode
continuous-variable systems, faithful representation of displacement
noise and strong squeezing requires very large photon-number
truncations, while the inclusion of non-Gaussian GKP states undergoing
repeated Gaussian interactions rapidly generates complex non-Gaussian
features that are numerically expensive to resolve. When extended to
many rounds of error correction, the explicit modeling of syndrome
measurements together with their measurement back-action leads to
proliferating state representations and accumulated truncation errors,
making long-time simulations dominated by numerical artifacts rather
than intrinsic physical behavior. As a result, circuit-level
simulations in this regime become both computationally prohibitive and
methodologically opaque, since simulation-specific errors can mask the
underlying noise-transformation processes. For the purposes of this
study—namely, understanding how correction layers reshape effective
displacement noise at the logical level, i.e., a reduced error-evolution
description provides a more controlled and physically transparent
framework that captures the essential noise dynamics without being
limited by large-scale Hilbert-space representations.

Three scenarios are considered:

\begin{enumerate}
\item {Without QEC:} The mode evolves under displacement noise without
  any error-correction mechanism.

\item {Only Gaussian error suppression:} A non-Gaussian correction
  step associated with GKP-type syndrome extraction is applied, which
  suppresses small Gaussian displacement errors.

\item {Concatenated code:} The GKP-assisted step is followed by an
  additional abstract correction layer representing the analog Steane
  structure, which corrects large displacement errors that cross
  lattice.
\end{enumerate}

In each QEC round, a displacement error (restricted to the position
quadrature in the simulation) is drawn from a Gaussian distribution
with variance $0.3$. Two error scenarios are considered.

(1) Gaussian-only errors, where no additional abrupt displacement is
introduced. This scenario provides a baseline comparison of the
error-suppression mechanisms under purely Gaussian errors.

(2) Gaussian + abrupt errors, where, in addition to the Gaussian
errors, an abrupt displacement error of fixed magnitude $+2\sqrt{\pi}$
is introduced once every 100 QEC rounds to model rare large
displacement events. This scenario is used to explicitly illustrate
the functionality of the outer analog Steane layer in correcting
lattice-crossing shifts.

For Gaussian displacement noise with variance $0.3$, the corresponding
lattice-crossing probability is approximately $10.6\%$, indicating
that such events are significant under this noise level. In all cases,
the results focus on the residual displacement $\zeta_{\mathrm{res}}$
after successive correction steps. For every scenario, a total of
$2,000$ Monte Carlo samples are generated to obtain the empirical
distribution of $\zeta_{\mathrm{res}}$. Figure~\ref{fig:simulation}
presents the evolution of the residual displacement for the six
scenarios. The blue solid curve denotes the mean residual displacement
in the current QEC round, $\zeta_{\mathrm{res,mean}}$, obtained by
averaging over all Monte Carlo samples. The black dashed curves
correspond to
$\zeta_{\mathrm{res,mean}} \pm \sigma_{\mathrm{current}}$, where
$\sigma_{\mathrm{current}}$ is the sample standard deviation in that
round. These quantities describe both the cumulative drift and the
statistical spread of the residual noise as the number of QEC rounds
increases. The value shown in the lower-left corner of each panel
indicates the empirical variance after the $1000^{\mathrm{th}}$ QEC
round. The first row corresponds to the Gaussian-only error scenario,
while the second row includes both Gaussian and abrupt displacement
errors. The first column represents the case without any QEC, the
second column corresponds to applying only the Gaussian-error
suppression circuit, and the third column shows the concatenated code.

Figure~\ref{fig:simulation} compares the evolution of residual
displacement noise under different correction strategies and noise
models. In the first column (no QEC), the residual displacement
exhibits a standard random-walk behavior. The variance grows
approximately linearly with the number of rounds and reaches
$\sigma^2 \sim 300$ after $1,000$ QEC rounds in both error
scenarios. This confirms that without error correction the
displacement noise accumulates. The second column shows the effect of
the Gaussian-error suppression circuit. In both error scenarios the
variance is reduced relative to the no-QEC case. However, the
reduction remains limited. Because the Gaussian error variance is
relatively large ($\sigma^2 = 0.3$ per round), the lattice-crossing
probability is significant ($\sim10.6\%$), and the suppression circuit
alone cannot efficiently correct these events. Consequently, the
observed variance reduction remains well above the theoretical $50\%$
suppression limit expected for purely Gaussian small-displacement
errors. The third column demonstrates the performance of the
concatenated code. In both the Gaussian-only and the
Gaussian-plus-abrupt noise scenarios, the residual variance is further
reduced and approaches the theoretical $50\%$ optimal suppression
factor. Notably, in the second row the abrupt displacement errors
introduce a systematic mean drift in the absence of an outer analog
Steane layer protection, which is clearly visible in the first two
columns. The concatenated code effectively removes this bias,
indicating that the outer analog Steane layer successfully corrects
lattice-crossing displacement shifts. Overall, the concatenated
architecture provides superior noise suppression in both stochastic
Gaussian error and mixed Gaussian-abrupt error environments. In
addition to reducing variance, it mitigates bias induced by rare large
displacement events, demonstrating robustness beyond Gaussian-noise
regimes.

In the context of bosonic quantum error correction, it is common to
inquire if a coding scheme admits a fault-tolerance threshold, i.e., a
physical noise level below which repeated QEC can, in principle, drive
the logical displacement noise arbitrarily close to zero.  However, as
established rigorously in~\cite{Haenggli2022NoThreshold}, no
oscillator-to-oscillator code operating under finite squeezing can
exhibit such a threshold.  Because the concatenated construction
studied here is subject to the same squeezing constraints, it likewise
does not possess a fundamental threshold, even though displacement
noise may be reduced within a limited operational regime.

We emphasize that the present concatenated oscillator-to-oscillator
architecture operates in a regime of strong error suppression rather
than threshold-based fault tolerance. For continuous-variable
encodings of logical information, it has been established that such
oscillator-to-oscillator codes do not admit a fault-tolerance
threshold. Even in the limit of ideal resources, residual displacement
noise cannot be driven arbitrarily close to zero. Consequently, over
sufficiently long computations, errors will inevitably accumulate.

This accumulation does not correspond to an abrupt failure
mechanism. Instead, the correction process reduces the rate of error
growth, so that for finite-depth or short-time computations the
residual errors remain well controlled and do not yet build up to a
level that compromises the encoded information. In this operational
regime, the concatenated code substantially extends the effective
coherence time and enables reliable logical processing.

The rate of residual error accumulation depends continuously on the
quality of the GKP resources. Increasing the squeezing strength of the
GKP states further suppresses the Gaussian component of the residual
displacement after each correction cycle, thereby slowing the
long-time diffusion of the logical displacement. This improvement is
gradual rather than threshold-like, consistent with the continuous
nature of the encoding.
Together, these considerations indicate that while
arbitrarily long fault-tolerant computation is not achievable in this
setting, the proposed architecture remains well suited for
finite-depth quantum computations with significantly enhanced
robustness.

\subsection{Comparison with qubit-based oscillator encoding}
\label{subsec:qubit_comparison}

An alternative approach to representing a bosonic oscillator is to
digitally encode its Hilbert space into multiple two-level systems.
In such qubit-based simulations, the oscillator's infinite-dimensional
Fock basis is truncated to a finite dimension \(N\), and each truncated
level \(|n\rangle\) is represented by a binary string over \(\log_2 N\)
qubits.  Importantly, the required truncation dimension \(N\) is not a
fixed constant. Rather, it must increase with the desired phase-space resolution
(how finely small displacements are resolved) and with the dynamic range
of excitations to be represented without truncation artifacts.  As a
result, the number of physical qubits and associated control overhead
grow as one seeks higher-precision oscillator dynamics.  This
resolution-dependent scaling constitutes a digitization cost intrinsic
to representing continuous degrees of freedom in a discrete Hilbert
space~\cite{Encinar2021PRA}.

The circuit complexity of DV simulations is also affected by this
truncation.  In the truncated Fock basis, canonical quadratures and
ladder operators correspond to dense \(N\times N\) matrices.  After
mapping to qubits, these operators generally expand into sums of many
nonlocal terms acting on multiple qubits.  Implementing Gaussian
unitaries in the DV picture therefore requires synthesizing evolutions
under such mapped Hamiltonians, typically via Trotterization or
related compilation techniques into multi-qubit gate sequences.
Consequently, the gate count and circuit depth increase with the
truncation dimension and target synthesis accuracy, even for basic
Gaussian transformations such as displacements, beam splitters, and
squeezers. This scaling is already substantial at moderate truncation
sizes.  For example, in the resource analysis of Ref.~\cite{Liu2025},
implementing a displacement operation on an all-qubit architecture
with a Fock-state cutoff of 64 levels requires a 7-qubit binary
register.  Using two Newton iterations to approximate the required
square root, which yields an arithmetic approximation error on the
order of $10^{-4}$, the total logical CNOT gate count is 36,788.
Furthermore, assuming that each logical qubit is likewise protected by
a Steane code, a comparative analysis is performed under the condition
that the logical qumode error rates are matched. Specifically, when
the logical qumode error rate is fixed at $17.89\%$ for both
architectures, the required number of physical quantum states,
physical gate count, circuit depth, and tolerable single-physical-gate
error bound on the CV platform. The qubit platforms are compared in
Table~\ref{tab:resource_comparison}. Under this matched logical error
rate, the CV platform requires substantially fewer physical quantum
states and significantly fewer physical gates, while also exhibiting a
markedly reduced circuit depth for implementing a single logical
displacement operation compared with the qubit platform. In addition,
for the same logical qumode error rate, the tolerable error bound per
physical gate in the CV platform is substantially higher than that in
the qubit platform, indicating less stringent requirements on physical
gate precision. These results demonstrate that, in terms of physical
quantum state count, physical gate count, circuit depth, and physical
gate precision requirements, the CV platform provides clear advantages
over the qubit platform.

\begin{table}
  \caption{Resource comparison between the CV system and the qubit
    system to simulate a single logical displacement operation under matched
    logical error rate of the entire circuit. For a CV system, the concatenated code
    proposed in this work is employed, whereas for a qubit system, the
    scheme reported in Ref.~\cite{Liu2025} is adopted.}
\label{tab:resource_comparison}
\begin{ruledtabular}
\begin{tabular}{lcc}
Metric & CV system & Qubit system \\
\hline
Logical qumode error rate & 17.89\% & 17.89\% \\
Number of quantum states & 15 & 49 \\
Logical gate count & 1 & 36{,}788 \\
Physical gate count & 15 & 257{,}516 \\
Circuit depth (max) & 3 & 36{,}788 \\
Single physical gate error bound & 2.53\% & 0.0000742\% \\
\end{tabular}
\end{ruledtabular}
\end{table}

This structural advantage originates from the fundamental encoding
paradigm of the CV architecture. The present CV-based encoding
directly employs a single physical oscillator as the logical carrier,
leveraging its native continuous Hilbert space.  Gaussian operations
such as displacement, squeezing, and beam splitting correspond to
elementary physical interactions (linear driving, parametric
amplification, and bilinear mode coupling) and are implemented as
native dynamical processes rather than compiled multi-qubit circuits.
The number of oscillators therefore does not scale with phase-space
resolution. Instead, the primary continuous-variable resource becomes
the achievable squeezing level (including the quality of approximate
GKP states), which determines the degree of Gaussian-noise suppression
without increasing the number of physical modes.

Overall, while qubit-based digital encodings offer a universal route to
simulating oscillators, the direct CV approach pursued here provides a
more hardware-efficient and physically transparent path toward
implementing oscillator-level quantum error correction.  By operating
directly in the native phase space of bosonic hardware and addressing
Gaussian displacement noise in its natural form, the CV framework
avoids the qubit overhead and truncation-induced circuit complexity
associated with DV simulations.

\section{Conclusion}
\label{sec:conclusion}

In this work, we have presented a concatenated CV quantum
error-correcting framework where a analog Steane code is combined with
GKP-based Gaussian error suppression.  The construction realizes
duality in the codes in that the inner GKP layer mitigates Gaussian
displacement noise, while the outer analog Steane code corrects
lattice-crossing and abrupt errors that occur beyond the suppression
capability of the GKP layer.  This duality corresponds to a separation
of error-mitigation roles in displacement space, where non-Gaussian
GKP resources suppress continuous Gaussian noise while the analog
Steane layer corrects discrete lattice-crossing events, enabling CV
error correction beyond Gaussian-only limits.  Most of our analytical
derivations have been conducted under idealized conditions with
infinitely squeezed states, establishing the theoretical upper limit
of the proposed architecture.

By analyzing the residual noise variance and the covariance structure
under finite squeezing, we demonstrated that the concatenated design
enables simultaneous suppression of Gaussian and abrupt displacement
errors, thus overcoming the Gaussian no-go constraint.
The results further show that the residual variance decreases
monotonically with the squeezing strength, indicating enhanced
precision in error localization and magnitude estimation.
Although a quantitative threshold for the GKP squeezing was not
specified, the outer analog Steane code effectively relaxes the resource
requirement for the inner GKP layer, suggesting reduced experimental
overhead.

Experimentally, the architecture requires only one non-Gaussian
resource, the GKP state, while all other operations remain Gaussian
and compatible with existing optical and superconducting CV platforms.
This feature highlights the feasibility of near-term experimental
demonstrations once approximate GKP states with sufficient fidelity
become available.

Future research will focus on exploring alternative non-Gaussian
resources for Gaussian-noise suppression and on developing hybrid
CV-DV concatenation schemes to further enhance robustness against
realistic noise in scalable quantum information processors.

\section*{Data Availability}
The code and data that support the findings of this study are publicly available on Zenodo at https://doi.org/10.5281/zenodo.19596790.

\begin{acknowledgments}
This work is supported by the U.S. Department of Energy, Office of Science, Advanced Scientific Computing Research, under contract number DE-SC0025384. This work was also funded in part by NSF grants MPS-2120757, NSF
PHY-2325080, OSI-2410675.
\end{acknowledgments}

\onecolumngrid
\vspace*{-1\baselineskip}
\appendix

\section{Derivation of the Gaussian Error Suppression Formula}
\label{app:derivation}
The probability density function of Eq.~(\ref{eq:xi_x_data_out}) can
be expressed as
\begin{equation}
\begin{aligned}
X\!\left(\xi_{x,\mathrm{data}}^{(\mathrm{out})}\right)
&= \frac{1}{2\pi\sigma^{2}}
\iint \delta\!\left(
\xi_{x,\mathrm{data}}^{(\mathrm{out})}
- \epsilon_{x,\mathrm{data}}
+ \tfrac{1}{2} R_{2\sqrt{\pi}}\!\left(\epsilon_{x,\mathrm{data}}+\epsilon_{x,\mathrm{GKP}}\right)
\right) \\
&\qquad \times
\exp\!\left[-\frac{\epsilon_{x,\mathrm{data}}^{2}+\epsilon_{x,\mathrm{GKP}}^{2}}{2\sigma^{2}}\right]
\, d\epsilon_{x,\mathrm{data}}\, d\epsilon_{x,\mathrm{GKP}} \\[4pt]
&= \sum_{m\in\mathbb{Z}}
\iint_{\;\epsilon_{x,\mathrm{data}}+\epsilon_{x,\mathrm{GKP}}\in
[\,2m\sqrt{\pi}-\sqrt{\pi},\, 2m\sqrt{\pi}+\sqrt{\pi}\,)}
\frac{1}{2\pi\sigma^{2}}\,
\delta\!\left(
\xi_{x,\mathrm{data}}^{(\mathrm{out})}
- \tfrac{1}{2}\epsilon_{x,\mathrm{data}}
+ \tfrac{1}{2}\epsilon_{x,\mathrm{GKP}}
- m\sqrt{\pi}
\right) \\
&\qquad \times
\exp\!\left[-\frac{\epsilon_{x,\mathrm{data}}^{2}+\epsilon_{x,\mathrm{GKP}}^{2}}{2\sigma^{2}}\right]
\, d\epsilon_{x,\mathrm{data}}\, d\epsilon_{x,\mathrm{GKP}} \\[4pt]
&= \frac{1}{\pi\sigma^{2}} \sum_{m\in\mathbb{Z}}
\int_{\,2m\sqrt{\pi}-\frac{\sqrt{\pi}}{2}-\xi_{x,\mathrm{data}}^{(\mathrm{out})}}^{\,2m\sqrt{\pi}+\frac{\sqrt{\pi}}{2}-\xi_{x,\mathrm{data}}^{(\mathrm{out})}} \\
&\qquad
\exp\!\left[
-\frac{\big(2\xi_{x,\mathrm{data}}^{(\mathrm{out})}+\epsilon_{x,\mathrm{GKP}}-2m\sqrt{\pi}\big)^{2}
+\epsilon_{x,\mathrm{GKP}}^{2}}{2\sigma^{2}}
\right]
\, d\epsilon_{x,\mathrm{GKP}} \\[4pt]
&= \frac{1}{\pi\sigma^{2}} \sum_{m\in\mathbb{Z}}
\exp\!\left[-\frac{\big(m\sqrt{\pi}-\xi_{x,\mathrm{data}}^{(\mathrm{out})}\big)^{2}}{\sigma^{2}}\right]
\int_{\,2m\sqrt{\pi}-\frac{\sqrt{\pi}}{2}-\xi_{x,\mathrm{data}}^{(\mathrm{out})}}^{\,2m\sqrt{\pi}+\frac{\sqrt{\pi}}{2}-\xi_{x,\mathrm{data}}^{(\mathrm{out})}} \\
&\qquad
\exp\!\left[
-\frac{\big(\epsilon_{x,\mathrm{GKP}}-(m\sqrt{\pi}-\xi_{x,\mathrm{data}}^{(\mathrm{out})})\big)^{2}}{\sigma^{2}}
\right]
\, d\epsilon_{x,\mathrm{GKP}} \\[4pt]
&= \frac{1}{2\sqrt{\pi}\,\sigma}
\sum_{m\in\mathbb{Z}}
\exp\!\left[-\frac{\big(m\sqrt{\pi}-\xi_{x,\mathrm{data}}^{(\mathrm{out})}\big)^{2}}{\sigma^{2}}\right]
\Bigg[\operatorname{erf}\!\left(\frac{m\sqrt{\pi}+\frac{\sqrt{\pi}}{2}}{\sigma}\right)
-\operatorname{erf}\!\left(\frac{m\sqrt{\pi}-\frac{\sqrt{\pi}}{2}}{\sigma}\right)\Bigg],
\end{aligned}
\label{eq:X_chain_detailed}
\end{equation}
where $\operatorname{erf}(x) = \frac{2}{\sqrt{\pi}}\!\int_0^x e^{-t^2}\,dt$ is the error function.

The variance of the corresponding probability density function is given by  
\begin{equation}
\begin{aligned}
\mathrm{Var}\!\left(\xi_{x,\mathrm{data}}^{(\mathrm{out})}\right)
&= \int \!\left(\xi_{x,\mathrm{data}}^{(\mathrm{out})}\right)^{2}
X\!\left(\xi_{x,\mathrm{data}}^{(\mathrm{out})}\right)
\, d\xi_{x,\mathrm{data}}^{(\mathrm{out})} \\
&= \frac{1}{2\sqrt{\pi}\,\sigma}
\sum_{m\in\mathbb{Z}}
\!\Bigg[
\operatorname{erf}\!\left(\frac{m\sqrt{\pi}+\frac{\sqrt{\pi}}{2}}{\sigma}\right)
-
\operatorname{erf}\!\left(\frac{m\sqrt{\pi}-\frac{\sqrt{\pi}}{2}}{\sigma}\right)
\Bigg] \\
&\qquad \times
\int_{-\infty}^{\infty}
\!\left(\xi_{x,\mathrm{data}}^{(\mathrm{out})}\right)^{2}
\exp\!\left[-\frac{\left(\xi_{x,\mathrm{data}}^{(\mathrm{out})}-m\sqrt{\pi}\right)^{2}}{\sigma^{2}}\right]
d\xi_{x,\mathrm{data}}^{(\mathrm{out})}.
\end{aligned}
\label{eq:var_x_data_detail}
\end{equation}
To further simplify the expression for 
$\mathrm{Var}\!\left(\xi_{x,\mathrm{data}}^{(\mathrm{out})}\right)$,  
we define
\begin{equation}
J_m = \int_{-\infty}^{\infty}
\!\left(\xi_{x,\mathrm{data}}^{(\mathrm{out})}\right)^{2}
\exp\!\left[-\frac{\left(\xi_{x,\mathrm{data}}^{(\mathrm{out})}-m\sqrt{\pi}\right)^{2}}{\sigma^{2}}\right]
d\xi_{x,\mathrm{data}}^{(\mathrm{out})}.
\label{eq:Jm_definition}
\end{equation}

Then,
\begin{align}
J_m
&= \int_{-\infty}^{\infty}
\left(\xi_{x,\mathrm{data}}^{(\mathrm{out})}-m\sqrt{\pi}\right)^{2}
\exp\!\left[-\frac{\left(\xi_{x,\mathrm{data}}^{(\mathrm{out})}-m\sqrt{\pi}\right)^{2}}{\sigma^{2}}\right]
d\xi_{x,\mathrm{data}}^{(\mathrm{out})} \notag\\
&\quad +\, 2m\sqrt{\pi}
\int_{-\infty}^{\infty}
\left(\xi_{x,\mathrm{data}}^{(\mathrm{out})}-m\sqrt{\pi}\right)
\exp\!\left[-\frac{\left(\xi_{x,\mathrm{data}}^{(\mathrm{out})}-m\sqrt{\pi}\right)^{2}}{\sigma^{2}}\right]
d\xi_{x,\mathrm{data}}^{(\mathrm{out})} \notag\\
&\quad +\, m^{2}\pi
\int_{-\infty}^{\infty}
\exp\!\left[-\frac{\left(\xi_{x,\mathrm{data}}^{(\mathrm{out})}-m\sqrt{\pi}\right)^{2}}{\sigma^{2}}\right]
d\xi_{x,\mathrm{data}}^{(\mathrm{out})} \notag\\
&= \sqrt{\pi}\,\sigma\!\left(m^{2}\pi+\frac{\sigma^{2}}{2}\right).
\label{eq:Jm_result_detail}
\end{align}

Therefore, we obtain  
\begin{equation}
\begin{aligned}
\mathrm{Var}\!\left(\xi_{x,\mathrm{data}}^{(\mathrm{out})}\right)
&= \frac{1}{2}
\sum_{m\in\mathbb{Z}}
\!\left(m^{2}\pi+\frac{\sigma^{2}}{2}\right) \\
&\quad\times
\Bigg[
\operatorname{erf}\!\left(\frac{m\sqrt{\pi}+\frac{\sqrt{\pi}}{2}}{\sigma}\right)
-
\operatorname{erf}\!\left(\frac{m\sqrt{\pi}-\frac{\sqrt{\pi}}{2}}{\sigma}\right)
\Bigg].
\end{aligned}
\label{eq:var_final_detail}
\end{equation}

\twocolumngrid

\nocite{*}

\bibliography{ref_new}

\end{document}